\documentclass[12pt]{article}

\usepackage{etoolbox}

\newtoggle{unnumberedsections}
\settoggle{unnumberedsections}{false}

\relax

\usepackage{amssymb,amsmath,amsthm}
\usepackage[colorlinks,citecolor=blue,urlcolor=blue]{hyperref}

\usepackage{cases}
\usepackage{graphicx}
\usepackage{epstopdf}
\usepackage{algorithm}
\usepackage[noend]{algpseudocode} 
\usepackage[comma]{natbib}
\usepackage{color}
\usepackage{xcolor}
\usepackage[dvipsnames]{xcolor}

\usepackage{subcaption} 

\usepackage[toc,page,title,titletoc,header]{appendix}
\usepackage{multirow}
\usepackage{booktabs}

\usepackage{setspace}
\usepackage{url}
 
\usepackage{cleveref}

\usepackage[normalem]{ulem}
 
\usepackage{xr}

\usepackage[top=2.7cm, bottom=3cm, left=2.7cm, right=2.7cm]{geometry}

\usepackage{mathtools}

\usepackage{sectsty}
\usepackage[subfigure]{tocloft}
\usepackage{titlesec}

\newtheorem{innerremark}{Remark}

\usepackage{amsmath}               
  {
      \theoremstyle{plain}

  }
\allowdisplaybreaks[1]

\usepackage{bbm}
\usepackage{bm}

\usepackage{authblk} 
\title{A generalized Bayesian approach to multiple changepoint analysis}

\date{\today}
\begin{document}
\author[1]{Yuhui Wang}  
\author[2]{Andrew M. Thomas}
\author[1]{Michael Jauch}

\affil[1]{Department of Statistics, Florida State University}
\affil[2]{Department of Statistics and Actuarial Science, University of Iowa}

    \maketitle
    
    \begin{abstract} We introduce a generalized Bayesian method for multiple changepoint analysis with a loss function inspired by multinomial logistic regression. The method does not require a specification of the data-generating process and avoids restrictive assumptions on the nature of changepoints. From the joint posterior distribution, we can make simultaneous inference on the locations of changepoints and the coefficients of a multinomial logistic regression model for distinguishing data across homogeneous segments. The multinomial logistic regression coefficients provide a familiar means of interpreting potentially complex changes. To select the number of changepoints, we leverage posterior summaries that measure whether the multinomial logistic classifier can distinguish data from either side of a potential changepoint. To simulate from the generalized posterior distribution, we present a Gibbs sampler based on P\'olya-Gamma data augmentation. We assess the accuracy and flexibility of our method through simulation studies featuring different types of changes and demonstrate its interpretability through applications to financial network data and topological data derived from nanoparticle videos.
    \end{abstract}

    
\section{Introduction}\label{Sec. Intro}

Data collected over time often exhibit abrupt structural changes. These changes can be complex, especially in modern data sets that feature high-dimensional measurements produced by complicated data-generating processes, and their number is typically unknown. The time indices that partition a heterogeneous series into homogeneous segments are called changepoints. Changepoint analysis is the process of determining the number, locations, and nature of these changepoints.
Since the pioneering work of \citet{page1954continuous} in the 1950s, changepoint analysis has been applied across numerous disciplines, including finance \citep{banerjee2020change}, bioinformatics \citep{hocking2013learning}, and climatology \citep{verbesselt2010detecting}. There is now a vast literature on methods for changepoint analysis. 
For a recent review of offline changepoint detection, our focus in the present article, we refer the reader to \citet{truong2020selective}.

The present article is motivated by three challenges that arise in changepoint analysis for modern data. First, it can be challenging to specify a model for the data within each homogeneous segment. Methods that assume data come from a parametric family (e.g., those described in \citealp{Chen2012}) are likely to be misspecified in practice, while nonparametric methods may be unwieldy or lack power to detect changes. It is difficult to strike the appropriate balance, especially with modern data types such as images or networks for which the data generating process may be quite elaborate. Second, it can be challenging, after identifying a changepoint, to interpret or characterize what exactly changed. This challenge is compounded when the data are higher-dimensional and changes are complex. Third, it can be challenging to properly account for uncertainty across related inferences in changepoint analysis (e.g., inferences regarding the number, locations, and nature of the changepoints). 

To address these challenges, we adopt a generalized Bayesian approach to multiple changepoint analysis, with a loss function inspired by multinomial logistic regression. Because the update from prior to posterior proceeds via a loss function rather than a likelihood function in the generalized Bayesian framework, we avoid the challenge of specifying a probability distribution for the data within each homogeneous segment. We can also avoid making restrictive assumptions regarding the nature of changepoints by choosing an appropriate representation for the data. From the joint posterior distribution, we can make simultaneous inference on the locations of changepoints and the coefficients of a multinomial logistic regression model for distinguishing data across homogeneous segments. The multinomial logistic regression coefficients provide a familiar means of interpreting potentially complex changes.

The proposed method generalizes the single changepoint method of \citet{thomas2024bayesian} to allow for simultaneous estimation, uncertainty quantification, and interpretation of multiple changepoints. A key component of our method is a novel approach to selecting the number of changepoints. In this approach, we fit our generalized Bayesian multiple changepoint method with a fixed but conservatively large number of changepoints and then determine how many of the fitted changepoints appear to correspond to true changepoints by examining posterior summaries. This approach provides the user with a posterior distribution of the number of fitted changepoints that appear to correspond to true changepoints. We also incorporate a horseshoe prior \citep{carvalho2010horseshoe} for the coefficients of the multinomial logistic regression, which can lead to more efficient and interpretable estimates when data are higher-dimensional and changes occur in a small number of coordinates.  To obtain posterior samples, we construct a Gibbs sampler via P\'olya-Gamma data augmentation \citep{polson2013bayesian}. Finally, we consider applications to complex data, including a video of a nanoparticle transitioning between ordered and disordered states and dynamic network data that records correlations between stock prices over time. 

Several recent articles have exploited classification techniques for changepoint detection. \citet{londschien2023random} introduced a non-parametric method based on a novel classifier log-likelihood ratio that leverages probabilistic predictions from a classifier such as a random forest. The ClaSP method \citep{ermshaus2023clasp} identifies change points by training a $k$-nearest neighbor classifier on windows before and after prospective changepoints, using cross-validation to construct a score profile whose peaks indicate potential changes. The method \texttt{changeAUC} of \citet{kanrar2025model} trains a classifier on held-out data from both ends of a series and then determines whether and where a changepoint exists by computing the area under the ROC curve (AUC) of this classifier on each potential split of the data. 
Our proposed generalized Bayesian method differs from these other methods by estimating multiple changepoints simultaneously, yielding more interpretable results, and providing uncertainty quantification regarding the changepoints and related parameters. 

The remainder of the article is structured as follows. Section~\ref{Sec. Methods} briefly reviews the most relevant background knowledge needed to understand the presentation of our proposed method in Section~\ref{Sec. bcmlr methods}. In Section~\ref{Sec. Simulation}, we compare our method against several alternative methods through simulation studies. Section~\ref{Sec. real data app} presents applications to real data. We conclude in Section~\ref{Sec. Discussion} with a summary of our contributions and a discussion of future work.
    \section{Background}\label{Sec. Methods}
We first review the generalized Bayesian framework of \citet{bissiri2016general} and the single changepoint method \texttt{BCLR} proposed by \citet{thomas2024bayesian}. 
    
\subsection{The generalized Bayesian framework}\label{Sec. generalized Bayes}

To make inferences regarding an unknown parameter $\theta$ given data $x$, a Bayesian approach uses a likelihood function $\mathcal{L}(\theta \mid x)$ to update a prior distribution with density function $\pi(\theta)$ to a posterior distribution with density function $\pi(\theta \mid x) \propto \mathcal{L}(\theta \mid x) \pi(\theta)$. To obtain the likelihood function, one must specify the data-generating process in the form of a probability distribution. As discussed in \citet{bissiri2016general}, there are scenarios in which this requirement can be prohibitively difficult or unnecessarily cumbersome. The data-generating process may be challenging to adequately specify, and the resulting likelihood function may be too complex to work with in practice. The parameter of interest may be low-dimensional, making inference via a fully specified model for the data a burden. Additionally, the parameter of interest may not directly index a family of density functions, presenting problems for the Bayesian approach. 
\citet{bissiri2016general} established that a valid, coherent update from $\pi(\theta)$ to $\pi(\theta \mid x)$ can be made for parameters which are connected to data $x$ via a loss function $\ell(\theta, x)$ rather than a likelihood function by setting
\begin{equation}\label{Eq. general-bayes-update} 
\pi(\theta \mid x) = \frac{\operatorname{exp}\left\{ - \ell(\theta, x)\right\} \pi(\theta)}{\int \operatorname{exp}\left\{ - \ell(\theta, x)\right\} \pi(\theta) \rm{d} \theta}.
\end{equation}

Developing a changepoint detection method under the generalized Bayesian framework will help to mitigate the challenges described in Section~\ref{Sec. Intro}. In particular, the generalized Bayesian approach uses a loss function rather than a likelihood, avoiding the need to specify the data-generating process for the data between changepoints. As we will see, the loss function can be chosen to offer an interpretable description of the potentially complex changes occurring at each changepoint. Finally, the generalized Bayesian approach yields a posterior distribution that can be used to quantify uncertainty regarding all unknown parameters, as in a standard Bayesian approach. 

The generalized Bayesian framework has been applied to other problems closely related to changepoint analysis. For example, \citet{rigon2023generalized} introduced a generalized Bayesian approach to probabilistic clustering. In a similar vein,  \citet{chakraborty2024gibbs} proposed a generalized Bayesian approach to fair clustering, providing uncertainty quantification about the optimal clustering configuration while ensuring protected attributes of subjects are balanced across all clusters. 
    \subsection{A generalized Bayesian single changepoint method}\label{Sec. bclr model}

We now review the single changepoint method proposed by \citet{thomas2024bayesian}, which we will refer to as \texttt{BCLR}. This method can be understood as a special case of the generalized Bayesian multiple changepoint method we will introduce in Section~\ref{Sec. bcmlr methods}. Our presentation is intended to make the parallels between these methods more evident. 

Let $\kappa \in \{1, \dots, N-1\}$ denote a single changepoint dividing a $p$-dimensional series $\bm{x}_1, \dots, \bm{x}_N$ into two homogeneous segments $\left\{1, \dots, \kappa \right\}$ 
and $\left\{\kappa+1, \dots, N \right\}.$ Let $\bm{X}$ be the $N \times p$ matrix whose rows correspond to $\bm{x}_1, \dots, \bm{x}_N.$ We can view \texttt{BCLR} as a generalized Bayesian method with the loss function
\begin{align} \label{bclr loss}
    \ell\left(\kappa, \bm{\beta}, \bm{X}\right) = - \log \left\{ \prod_{i=1}^{\kappa} \frac{1}{1+e^{\bm{x}_i^{\top}\bm{\beta}}} \prod_{i=\kappa+1}^N \frac{e^{\bm{x}_i^{\top} \bm{\beta}}}{1+e^{\bm{x}_i^{\top}\bm{\beta}}}\right\}
\end{align}
connecting the parameters of interest $\bm{\theta} = (\kappa, \bm{\beta})$ to the data $\bm{X}$. This loss function is closely related to logistic regression: If the changepoint $\kappa$ were known, then \eqref{bclr loss} would reduce to the negative log-likelihood function of a logistic regression model for distinguishing between pre- and post-changepoint data. The regression coefficients allow one to interpret a changepoint just as one would interpret the coefficients of a logistic regression. The generalized posterior density satisfies
\begin{align} \label{bclr posterior}
    \pi \left(\kappa, \bm{\beta} \mid \bm{x}_1, \dots, \bm{x}_N\right) \propto \operatorname{exp}\left\{- \ell\left(\kappa, \bm{\beta}, \bm{X}\right)\right\} \pi\left(\kappa, \bm{\beta} \right)
\end{align} where $\pi\left(\kappa, \bm{\beta} \right)$ is the prior density. As a default setting, \citet{thomas2024bayesian} chose a uniform prior for the changepoint and a multivariate Gaussian prior for the regression coefficients, with the two parameters being independent \textit{a priori}. 

\section{Multiple changepoint detection} 
\label{Sec. bcmlr methods}

Most real world applications of changepoint analysis feature multiple changepoints. Multiple changepoint methods are often constructed by repeatedly applying a single changepoint method. Common approaches include binary segmentation and bottom-up segmentation \citep{vostrikova1981detecting, fryzlewicz2014wild, zhang2021graph, ross2021nonparametric, korkas2022ensemble, kovacs2023seeded}. Binary segmentation applies a single changepoint method recursively by splitting a series into two segments if a changepoint is found, and then applying the same method to each of the resulting segments. This process continues until no more changepoints are found or a minimum segment length is reached. Bottom-up segmentation begins with a large number of potential changepoints dividing a series into numerous segments and then merges neighboring segments if there is no evident difference between them. These approaches can be convenient and effective, but they are not ideal for generalizing the single changepoint approach of \citet{thomas2024bayesian} because important advantages of that method, such as its interpretability and coherent uncertainty quantification, would be lost in the process. Instead, in this section, we present a generalized Bayesian multiple changepoint method with a loss function inspired by multinomial logistic regression that retains the advantages of \texttt{BCLR} by providing simultaneous inference for multiple changepoints and associated parameters.  An important component of our method is a novel approach to selecting the number of changepoints based on posterior summaries. We also consider a horseshoe prior for the coefficients of the multinomial logistic regression, which can lead to more efficient and interpretable estimates when data are higher-dimensional and changes occur in a small number of coordinates. 

\subsection{A generalized Bayesian multiple changepoint method} \label{Sec. bcmlr model}

We now suppose there are $L$ changepoints $\kappa_1, \dots, \kappa_L \in  \{1, \dots, N-1\}$ with $\kappa_1 < \dots < \kappa_L$ that divide the series $\bm{x}_1, \dots, \bm{x}_N$ into $J = L+1$ segments. Let $\bm{\kappa} = (\kappa_1, \dots, \kappa_L)^{\top}$ and, for notational convenience, set $\kappa_0 = 0$ and  $\kappa_{L+1} = N$. For each $j \in \{1, \dots, L+1\}$, the two consecutive changepoints $\kappa_{j-1}$ and $\kappa_j$ define the $j$th homogeneous segment $\left\{\kappa_{j-1} +1, \dots, \kappa_{j}\right\}$. 
Again, let $\bm{X}$ be the $N \times p$ matrix whose rows correspond to $\bm{x}_1, \dots, \bm{x}_N.$ 

Under the generalized Bayesian framework, we choose the loss function 
\begin{align}\label{Eq. bcmlr loss}
    \ell\left(\bm{\kappa}, \bm{\beta}_1, \dots, \bm{\beta}_J, \bm{X}\right) = - \log \left\{\prod_{i = 1}^{\kappa_1} q_{i1} \dots \prod_{i = \kappa_L +1}^{N} q_{iJ} \right\}, \text{ where } q_{ij} = \frac{e^{\bm{x}_i^{\top}\bm{\beta}_j}}{\sum_{k=1}^J e^{\bm{x}_i^{\top}\bm{\beta}_k}},
\end{align}
to connect the parameters of interest $\bm{\theta} = \left(\bm{\kappa}, \bm{\beta}_1, \dots, \bm{\beta}_J\right)$ to the data $\bm{X}$. This loss function is closely related to multinomial logistic regression. In particular, if the changepoints $\kappa_1, \dots, \kappa_L$ were known, then \eqref{Eq. bcmlr loss} would reduce to the negative log-likelihood function of a multinomial logistic regression model for distinguishing data across homogeneous segments. In the usual scenario when $\kappa_1, \dots, \kappa_L$ and $\bm{\beta}_1, \dots, \bm{\beta}_J$ are unknown, minimizing the loss function would correspond to finding the combination of $\kappa_1, \dots, \kappa_L$ and $\bm{\beta}_1, \dots, \bm{\beta}_J$ such that the multinomial logistic regression model with coefficients $\bm{\beta}_1, \dots, \bm{\beta}_J$ best discriminates data across homogeneous segments. The regression coefficients can be interpreted just as we would interpret
the coefficients of a multinomial logistic regression, providing a familiar way of understanding potentially complex changes. 

The generalized Bayesian posterior density satisfies 
\begin{align} \label{Eq. bcmlr posterior}
     \pi \left(\bm{\kappa}, \bm{\beta}_1, \dots, \bm{\beta}_J | \bm{X}\right) \propto \operatorname{exp}\left\{ - \ell\left(\bm{\kappa}, \bm{\beta}_1, \dots, \bm{\beta}_J, \bm{X} \right)\right\} \pi\left(\bm{\kappa}, \bm{\beta}_1, \dots, \bm{\beta}_J\right).
\end{align}
Because $\operatorname{exp}\left\{ - \ell\left(\bm{\kappa}, \bm{\beta}_1, \dots, \bm{\beta}_J, \bm{X} \right)\right\}$ is bounded above by one, a proper prior distribution leads to a proper posterior distribution. By default, we let $\pi\left(\bm{\kappa}, \bm{\beta}_1, \dots, \bm{\beta}_J\right) = \pi\left(\bm{\kappa})\pi( \bm{\beta}_1, \dots, \bm{\beta}_J\right)$ 
with
\begin{align}\label{Eq. bcmlr kappa prior}
\pi\left(\bm{\kappa}\right) \propto \left(\frac{1}{\kappa_1 - \kappa_0}\right)^{\kappa_1-\kappa_0} \left(\frac{1}{\kappa_2 - \kappa_1}\right)^{\kappa_2-\kappa_1}  \dots \left(\frac{1}{\kappa_{L+1} - \kappa_{L}}\right)^{\kappa_{L+1}-\kappa_L} .
\end{align}
This prior for $\bm{\kappa}$ penalizes small segments and assigns the greatest mass to the case when all homogeneous segments are of equal length.

We consider two different prior specifications for the regression coefficients. Before describing these, we first remark that, as in standard multinomial logistic regression, we must choose a reference class and set the corresponding coefficient vector to zero. We let class $J$ be the reference class and set $\bm{\beta}_J = \bm{0}$.  When $p$ is small, we assign independent multivariate Gaussian priors to each of the $J-1$ remaining coefficient vectors, with $\bm{\beta}_j \sim N \left(\bm{m}_{0j}, \bm{V}_{0j} \right)$ for each $j \in \left\{1, \dots, J-1\right\}.$ When $p$ is large and we expect changes to occur in only a few dimensions, we adopt a horseshoe prior \citep{carvalho2010horseshoe} for $\bm{\beta}_1, \ldots, \bm{\beta}_{J-1}.$ The horseshoe prior has a Gaussian scale mixture representation, with 
\begin{align*}
\beta_{dj} \mid \lambda_{dj}, \tau \sim N\left(0, \lambda_{dj}^2 \tau^2\right), \quad 
\lambda_{dj} \sim  C^{+}(0,1), \quad 
\tau \sim C^{+}(0,1)
\end{align*}
for $d = 1, \dots, p$ and $j=1, \dots, J-1$. Here, $C^+(0,1)$ denotes the half-Cauchy distribution with a location parameter of zero and a scale parameter of one. Each $\lambda_{dj}$ is a local scale parameter, while $\tau$ is the global scale parameter.

Since its introduction, the horseshoe prior has been widely adopted as a prior for sparse parameters because of its desirable statistical properties and its computational tractability. The horseshoe prior is meant to aggressively shrink estimates toward zero when the true coefficient is zero, but avoid overshrinking estimates of non-zero coefficients. This leads to more efficient and interpretable estimates of sparse parameters. \citet{makalic2015simple} introduced a simple Gibbs sampler for posterior inference with the horseshoe prior, leveraging the fact that the half-Cauchy distribution can be expressed as a scale mixture of inverse gamma distributions to arrive at a data-augmented representation of the horseshoe prior. In our context, the data-augmented representation is 
\begin{align} \label{Eq. HS prior}
    \beta_{dj} \mid  \lambda_{dj}, \tau  \sim N\left(0, \lambda_{dj}^2 \tau^2\right), 
    \quad 
    \lambda_{dj}^2 \mid \nu_{dj} \sim \operatorname{IG}\left(\frac{1}{2}, \frac{1}{\nu_{dj}}\right), 
    \quad  
    \tau^2 \mid \xi \sim \operatorname{IG} \left(\frac{1}{2}, \frac{1}{\xi}\right)
\end{align}
with $\nu_{dj} \sim \operatorname{IG}\left(\frac{1}{2}, 1\right)$ for $d = 1, \dots, p$ and $j=1, \dots, J-1,$ and $\xi \sim \operatorname{IG} \left(\frac{1}{2}, 1\right).$ 
This representation leads to a Gibbs sampler with inverse gamma full conditional distributions for the local and global scale parameters as well as the newly-introduced auxiliary variables.

We will refer to this method as $\texttt{bcmlr},$ which is a stylized acronym for ``generalized \textbf{B}ayesian \textbf{c}hangepoint detection with a loss inspired by \textbf{m}ultinomial \textbf{l}ogistic \textbf{r}egression." Section \ref{Sec. select num of CPs} will describe an approach to selecting the number of changepoints. As with the \texttt{BCLR} method of \citet{thomas2024bayesian}, we omit intercept parameters and center the data. We also standardize the data to facilitate the use of a default prior for the regression coefficients when data have different scales.

\subsection{A Gibbs sampler for multiple changepoint detection} 
\label{Sec. bcmlr Gibbs sampler}

To simulate from the posterior distribution, we apply Pólya-Gamma data augmentation to obtain an augmented posterior density that yields tractable full conditional distributions. Following the Section 6.3 of the supplement of \citet{polson2013bayesian}, which discusses Pólya-Gamma data augmentation for multinomial logistic regression, we introduce a minor change of notation, re-expressing the loss function \eqref{Eq. bcmlr loss} as 
$$
    \ell\left(\bm{y}, \bm{\beta}_1, \dots, \bm{\beta}_J \mid \bm{X}\right) 
    = - \log \left\{\prod_{i=1}^N \prod_{j=1}^J \frac{e^{\eta_{ij}y_{ij}}}{1 + e^{\eta_{ij}}} \right\} 
$$
where $\eta_{ij} \left(\bm{x}_i, \bm{\beta}_j\right)  = \bm{x}_i^{\top} \bm{\beta}_j - \log \sum_{k \neq j} e^{\bm{x}_i^{\top} \bm{\beta}_k}$ with $y_{ij} = 1$ if $\kappa_{j-1} < i \leq \kappa_{j}$ and $y_{ij} = 0$ otherwise. This alternative expression for the loss, introduced in \citet{held2006bayesian}, resembles the loss function in binomial logistic regression and allows us to leverage Pólya-Gamma data augmentation to obtain the posterior density of $(\bm{\kappa}, \bm{\beta}_1, \dots, \bm{\beta}_J, \bm{\omega}),$ where $\bm{\omega}$ is an $N \times J$ matrix of Pólya-Gamma auxiliary variables. From this augmented posterior density, we can derive the full conditional distributions necessary to construct a Gibbs sampler. 

We will present two versions of the Gibbs sampler that differ according to the prior distribution assigned to the regression coefficients. In both versions, the full conditional distributions of the changepoints and the Pólya-Gamma auxiliary variables are the same. 
For $l \in \{1, \dots, L\}$, let $\bm{\kappa}_{-l}$ denote the collection of all changepoints except for $\kappa_l$. The full conditional distribution of $\kappa_l$ is a discrete distribution that satisfies 
\begin{align}\label{Eq. bcmlr kappa_l full-conditional}
    &\pi\left(\kappa_l \mid \bm{\kappa}_{-l}, \bm{\beta}_1, \dots, \bm{\beta}_J, \bm{X}\right) \propto \pi\left(\bm{\kappa}\right) \prod_{i=\kappa_{l-1}+1}^{\kappa_l} q_{il} \prod_{i = \kappa_l+1}^{\kappa_{l+1}} q_{i,l+1}
\end{align}
on the support $\left\{\kappa_{l-1}+1, \dots, \kappa_{l+1}-1\right\}$. The full conditional distribution of $\bm{\omega}$ is the same as in \citet{polson2013bayesian}: 
\begin{equation} \label{Eq. full condi omega}
    \omega_{ij} \mid \bm{\beta}_1, \dots, \bm{\beta}_{J},  \bm{x}_i \sim \operatorname{PG}\left(1, \eta_{ij}\right)
\end{equation}
for $i \in \{1, \dots, N\}$ and $j \in \{1, \dots, J-1\}$. 

We first present the Gibbs sampler based on Gaussian priors for $\bm{\beta}_1, \dots, \bm{\beta}_{J-1}$. For $j \in \{1,\dots,J-1\}$, we denote by $\bm{\beta}_{-j}$ the collection of the remaining $J-2$ coefficient vectors excluding $\bm{\beta}_j$. When $\bm{\beta}_j$ has a Gaussian prior $N(\bm{m}_{0j}, \bm{V}_{0j})$, the full conditional distribution of $\bm{\beta}_j$ is 
\begin{align}
    \bm{\beta}_j \mid \bm{\beta}_{-j}, \bm{\omega}_j, \bm{\kappa}, \bm{X} &\sim N(\bm{m}_j, \bm{V}_j), \nonumber\\
    \bm{V}_j = (\bm{X}^{\top}\bm{\Omega}_j \bm{X}+\bm{V}_{0j}^{-1})^{-1}, \quad \bm{m}_j &= \bm{V}_j\left(\bm{X}^{\top}(\bm{\Omega}_j \bm{c}_j + \bm{\delta}_j) + \bm{V}_{0j}^{-1} \bm{m}_{0j}\right), \nonumber\\
    \bm{\Omega}_j = \operatorname{diag}(\bm{\omega}_j), \quad \bm{\omega}_j &= \left(\omega_{1j}, \dots, \omega_{Nj}\right)^{\top}, \label{Eq. Omega_j}\\
    \bm{\delta}_j = \bm{y}_j - \mathbbm{1}_N  \cdot \frac{1}{2}, \quad \bm{y}_j &= \left(y_{1j}, \dots, y_{Nj}\right)^{\top}\label{Eq. delta_j} \\
    c_{ij} = \operatorname{log}\sum_{k\neq j} \operatorname{exp}\left( \bm{x}_i^{\top} \bm{\beta}_k\right), \quad \bm{c}_j &= \left(c_{1j}, \dots, c_{Nj}\right)^{\top}, \label{Eq. c_ij}
\end{align} where $j \in \{1, \dots, J-1\}$ and $i\in\{1,\dots,N\}$. 
With Gaussian priors for the regression coefficients and the discrete prior distribution \eqref{Eq. bcmlr kappa prior} for the changepoints, the Gibbs sampler targeting the Pólya-Gamma augmented posterior distribution iterates through the following three steps: 
$$
\begin{aligned}
    \kappa_l \mid \bm{\kappa}_{-l}, \bm{\beta}_1, \dots, \bm{\beta}_{J}, \bm{X} &\sim \pi\left(\kappa_l \mid \bm{\kappa}_{-l}, \bm{\beta}_1, \dots, \bm{\beta}_{J}, \bm{X}\right), \\
    \omega_{ij} \mid \bm{\beta}_j, \bm{x}_i
    &\sim \operatorname{PG}\left(1, \eta_{ij}\right),\\
    \bm{\beta}_j \mid \bm{\beta}_{-j}, \bm{\omega}_j, \bm{\kappa}, \bm{X} &\sim N(\bm{m}_j, \bm{V}_j), 
\end{aligned}
$$
where $ l \in \{ 1, \dots, L\}, i \in\{ 1, \dots, N \}$ , and $j \in \{1, \dots, J-1\}$.

Next, we describe the Gibbs sampler that results from assigning the horseshoe prior described in Section~\ref{Sec. bcmlr model} to the regression coefficients $\bm{\beta}_1, \dots, \bm{\beta}_{J-1}.$ The data-augmented representation of the horseshoe prior in \eqref{Eq. HS prior} involves additional auxiliary variables that must be updated while Gibbs sampling. The full conditional distribution of $\bm{\beta_j}$ becomes
$$
\begin{aligned}
    \bm{\beta}_j \mid \bm{\beta}_{-j}, \bm{\omega}_j, \bm{\kappa}, \bm{X}, \bm{\lambda}_j, \bm{\nu}_j, \tau, \xi &\sim N(\bm{m}_j, \bm{V}_j), \\
    \bm{V}_j = (\bm{X}^{\top}\bm{\Omega}_j \bm{X}+\bm{\Sigma}_{0j})^{-1}, \quad
    \bm{m}_j &= \bm{V}_j\left(\bm{X}^{\top}(\bm{\Omega}_j \bm{c}_j + \bm{\delta}_j)\right),\\
    \bm{\Sigma}_{0j} = \operatorname{diag}\left(\lambda_{1j}^2\tau^2, \dots, \lambda_{pj}^2\tau^2\right), \quad \bm{\lambda}_j &= \left(\lambda_{1j}, \dots, \lambda_{pj}\right)^{\top}, \quad \bm{\nu}_j = \left(\nu_{1j}, \dots, \nu_{pj}\right)^{\top}, 
\end{aligned}
$$
where $\bm{\Omega}_j$, $\bm{c}_j$ and $\bm{\delta}_j$ remain the same as in \eqref{Eq. Omega_j}-\eqref{Eq. c_ij}. The full conditional distributions of $\lambda_{dj}, \nu_{dj}$, $\tau$, and $\xi$ are available in \citet{bhattacharyya2022applications}. 
Putting this all together, with the horseshoe prior \eqref{Eq. HS prior} for the regression coefficients and the discrete prior distribution \eqref{Eq. bcmlr kappa prior} for the changepoints, the Gibbs sampler targeting the augmented posterior distribution iterates through the following steps: 
\begin{align*}
    \kappa_l \mid \bm{\kappa}_{-l}, \bm{\beta}_1, \dots, \bm{\beta}_{J}, \bm{X} &\sim \pi\left(\kappa_l \mid \bm{\kappa}_{-l}, \bm{\beta}_1, \dots, \bm{\beta}_{J}, \bm{X}\right), \\ 
    \omega_{ij} \mid \bm{\beta}_j, \bm{x}_i
    &\sim \operatorname{PG}\left(1, \eta_{ij}\right), \\ 
    \bm{\beta}_j \mid \bm{\beta}_{-j}, \bm{\omega}_j, \bm{\kappa}, \bm{X}, \bm{\lambda}_j, \bm{\nu}_j, \tau, \xi &\sim N(\bm{m}_j, \bm{V}_j), \\ 
    \lambda_{dj}^2 \mid \nu_{dj}, \beta_{dj}, \xi, \lambda_{dj}^2 &\sim \operatorname{IG}\left(1, \frac{1}{\nu_{dj}} + \frac{\beta_{dj}^2}{2\tau^2}\right), \\
    \nu_{dj} \mid \lambda_{dj}^2, \beta_{dj}, \xi, \tau^2 &\sim \operatorname{IG}\left(1, 1+ \frac{1}{\lambda_{dj}^2}\right) \\
    \tau^2 \mid \nu_{dj}, \beta_{dj}, \xi, \lambda_{dj}^2 &\sim \operatorname{IG}\left(\frac{p+1}{2}, \frac{1}{\xi} + \sum_{d=1}^p \frac{\beta_{dj}^2}{2\lambda_{dj}^2}\right), \\
    \xi \mid \tau^2 &\sim \operatorname{IG}\left(1, 1+\frac{1}{\tau^2}\right), 
\end{align*}
where $ l \in \{ 1, \dots, L\}, i \in\{ 1, \dots, N \}$ , and $j \in \{1, \dots, J-1\}$. 

Software implementing both Gibbs samplers is available in the first author's GitHub repository \footnote{\texttt{https://github.com/YuhuiChloe/bcmlr}}. In the implementation, we employ the strategy described in \citet{bhattacharya2016fast} when updating the regression coefficients. As is common practice in the changepoint detection literature, we impose a minimum segment length $m$ between consecutive changepoints, which modifies the support of the full conditional distribution \eqref{Eq. bcmlr kappa_l full-conditional}. For simplicity, we exclude the minimum segment length constraint in \eqref{Eq. bcmlr kappa_l full-conditional}. 

The horseshoe prior can also be advantageous in the context of the single changepoint method \texttt{bclr} reviewed in Section~\ref{Sec. bclr model}. We present a simplified version of the Gibbs sampler resulting from the horseshoe prior, specific to the single changepoint setting, in 
the Supplementary Material.

\subsection{Initialization and mixing} \label{Sec. Accelerating mixing} 

We have found that initializing the Gibbs sampler with evenly spaced changepoints works well in most scenarios, and this initialization scheme was used in the simulation studies of Section~\ref{Sec. Simulation} as well as the real data applications of Section~\ref{Sec. real data app}. For some very challenging configurations of the true changepoints, however, the Gibbs sampler may suffer from poor mixing (e.g., the Markov chain may get stuck in a local mode and fail to explore the larger space of changepoint configurations). To address this problem, we considered two alternative strategies: (1) initializing the Gibbs sampler at changepoint locations estimated by a fast external method and (2) applying parallel tempering \citep{geyer1991markov}. Both of these strategies are available in our software implementation of \texttt{bcmlr}. In particular, the code supports initialization via \texttt{kcp} \citep{arlot2019kernel}, \texttt{e.divisive} \citep{matteson2014nonparametric}, and \texttt{MultiRank} \citep{lung2015homogeneity}. Users can also apply the non-reversible parallel tempering method of  \citet{syed2022non} as described in 
the supplementary material. In our experiments, initialization via a fast external method did not consistently outperform even initialization, and the occasional benefits of parallel tempering did not outweigh the additional computational burden. Thus, we recommend even initialization as the default choice.

\subsection{Determination of the number of changepoints} 
\label{Sec. select num of CPs}

Thus far, we have assumed the number of changepoints is known. While this is sometimes the case, the number of changepoints $L_{\text{true}}$ is typically unknown in practice. A natural Bayesian approach would be to assign a prior to the number of changepoints and then simulate from the resulting posterior distribution using reversible jump Markov chain Monte Carlo \citep{green1995reversible}. However, this approach would incur a substantial computational cost. We present a computationally tractable alternative that involves applying the \texttt{bcmlr} method with a conservative upper bound $L_{\text{fitted}}$ for the number of changepoints and then determining how many of the fitted changepoints correspond to true changepoints by examining posterior summaries.

We take the perspective that a fitted changepoint corresponds to a true changepoint if the multinomial logistic regression obtained from our generalized posterior distribution can classify data on either side of the fitted changepoint better than a coin flip. From this perspective, it is natural to consider whether the area under the ROC curve (AUC) \citep{peterson1954theory} of the multinomial logistic regression applied to this binary classification task is greater than 0.5, because an AUC less than or equal to $0.5$ indicates that a classifier is no better than a coin flip. 

More concretely, our approach proceeds as follows. We run the Gibbs sampler in Section~\ref{Sec. bcmlr Gibbs sampler} to obtain fitted changepoints $\kappa_1^{(s)},\dots, \kappa_{L_{\text{fitted}}}^{(s)}$ from the posterior density \eqref{Eq. bcmlr posterior}. Here, $s\in \{1, \dots, S\}$ denotes the iteration number. For each iteration $s\in \{1, \dots, S\}$ and changepoint $l \in \left\{1, \dots, L_{\text{fitted}}\right\}$, we compute an AUC based on data in the two neighboring segments on either side of $\kappa_l$ using class labels $y_{i, l+1}^{(s)}$ and probabilities $\tilde{q}_{i,l+1}^{(s)}$. The class labels are binary with $y_{i,l+1}^{(s)} = 0$ for $i \in \left\{\kappa_{l-1}^{(s)}+1, \dots, \kappa_{l}^{(s)}\right\}$ and $y_{i, l+1}^{(s)} = 1$ for $i \in \left\{\kappa_{l}^{(s)}+1, \dots, \kappa_{l+1}^{(s)}\right\}$. The probability $\tilde{q}_{i, l+1}^{(s)}$ is defined as $\tilde{q}_{i, l+1}^{(s)} = q_{i, l+1}^{(s)}/\left(q_{i, l+1}^{(s)} + q_{i,l}^{(s)}\right)$ with $q_{i, l+1}^{(s)}$ and $q_{i,l}^{(s)}$ given in \eqref{Eq. bcmlr loss}. This probability can be understood as the conditional probability (according to the multinomial logistic regression) of $\bm{x}_i$ belonging to class $l+1$ given that $\bm{x}_i$ belongs to either class $l$ or class $l+1$. If the AUC is above the threshold $\tau = 0.5$, we consider $\kappa_l^{(s)}$ a true changepoint and set an indicator $R_l^{(s)}$ equal to 1; otherwise, we do not consider $\kappa_l^{(s)}$ a true changepoint and set $R_l^{(s)} = 0$. The sum of the indicators $L_{\text{true}}^{(s)} = \sum_{l=1}^{L_{\text{fitted}}} R_l^{(s)}$ is the number of fitted changepoints that correspond to true changepoints in the sense described above at iteration $s.$ In the end, $L_{\text{true}}^{(1)}, \dots, L_{\text{true}}^{(S)}$ approximate the posterior distribution of the number of fitted changepoints that correspond to true changepoints. We take the posterior mode of $L_\text{true}$ as our point estimate $\hat{L}_\text{true}.$

In practice, we recommend a few modifications to the approach described above. Since the AUC estimate calculated above is based on a potentially small number of observations in the two neighboring segments, we recommend computing a $(1-\alpha) \times  100$\% confidence interval for the AUC and using the lower bound of this interval instead of the AUC estimate itself. We use the \texttt{pROC} package \citep{robin2011proc} to compute AUC estimates and  confidence intervals. To avoid selecting too many changepoints due to overfitting, we recommend holding out  every $\zeta$th observation to compute these confidence intervals while using the remaining data to inform the posterior distribution of the $L_{\text{fitted}}$ changepoints. One must choose a minimum segment length $m$ greater than $\zeta$ to ensure that every segment includes at least one held-out sample. After obtaining an estimate $\hat{L}_{\text{true}}$ of the number of changepoints, we recommend reapplying the \texttt{bcmlr} method to the entire series with $L = \hat{L}_{\text{true}}.$  This will enable easier interpretation of the regression coefficients and allow the changepoints to occur at any location in the series. If there is substantial uncertainty regarding the number of fitted changepoints that correspond to true changepoints (as in, for example, Figure~\ref{fig:DJIA-numCPdistribution}), the user may want to refit the \texttt{bcmlr} method for each of the most probable values of $L.$ We provide a function to select the number of changepoints in the GitHub repository. In that function, all of the modifications recommended above are implemented by default.

The approach described above includes two interpretable tuning parameters. The tuning parameter $\alpha$ (with a default value of $\alpha = 0.05$) reflects the required level of certainty to determine that a change has occurred. The tuning parameter $\tau$ (with a default value of $\tau = 0.5$) determines the magnitude of a change necessary for the change to be detected. 

    \section{Simulation studies}\label{Sec. Simulation}

We compare the proposed method (with the horseshoe prior) to five non-parameteric methods: \texttt{e.divisive} \citep{matteson2014nonparametric}, \texttt{kcp} \citep{arlot2019kernel}, \texttt{MultiRank} \citep{lung2015homogeneity}, \texttt{ClaSP} \citep{ermshaus2023clasp}, and \texttt{changeforest} \citep{londschien2023random}. The latter two methods, which leverage classifiers for changepoint detection, were reviewed in Section~\ref{Sec. Intro} along with \texttt{changeAUC} \citep{kanrar2025model}. Although \citet{kanrar2025model} described an extension of their \texttt{changeAUC} method to the multiple changepoint setting, we found that this extension (with the default number of permutations) took much longer to run than \texttt{bcmlr} and the five non-parametric methods. For this reason, we do not include \texttt{changeAUC} in our simulation studies. The \texttt{e.divisive} method combines bisection \citep{vostrikova1981detecting} with a multivariate divergence measure \citep{szekely2005hierarchical} to recursively identify changepoints. The kernel-based \texttt{kcp} method detects changepoints for data that can be equipped with a positive semidefinite kernel. The \texttt{MultiRank} method identifies segment boundaries by optimizing a homogeneity test statistic. Implementation details, including tuning parameters and available packages, for all methods in the simulation studies can be found in 
the Supplementary Material. 

We use the adjusted Rand index \citep{morey1984measurement} to assess the accuracy of a changepoint detection method. Given a partition of the data based on the true changepoints and a partition of the data based on estimated changepoints, the Rand index \citep{rand1971objective} is equal to the proportion of the $\binom{N}{2}$ pairs of observations that are either in the same subset in both partitions or in different subsets in both partitions. The adjusted Rand index (ARI) is the normalized difference between the Rand index and the expected Rand index for a random partition. An ARI of 1 indicates perfect changepoint estimation. 

\subsection{Simulation scenarios}
We evaluate our method and the competing methods in 12 scenarios that differ based on the type of changes (change in mean, change in covariance, or change in both mean and covariance), whether the number of changepoints is known or estimated, and the dimensionality of the data (lower-dimensional vs. higher-dimensional). 
In each scenario, our evaluation is based on 100 simulated series. Each series has length $N = 600$ and contains two true changepoints, $\kappa_1 = 100$ and $\kappa_2 = 500$. The data-generating processes are described below.

\paragraph{Changes in mean (CIM):} For $1\leq i \leq \kappa_1$ and $\kappa_2 +1 \leq i \leq N$, we generate $\bm{x}_i$ from the multivariate Gaussian distribution $N\left(\bm{0}, \bm{I}\right)$ where $\bm{0}$ is a $p$-dimensional vector of zeros and $\bm{I}$ is the $p\times p$ identity matrix. For $\kappa_1 +1 \leq i \leq \kappa_2$, we generate $\bm{x}_i$ from the multivariate Gaussian distribution $N\left(\bm{\mu}, \bm{I}\right)$ where the first two entries in the mean vector $\bm{\mu}$ equal 2, the last two entries equal $-2$, and the $p-4$ entries in the middle equal 0. We set $p = 14$ in the lower-dimensional scenario and $p = 40$ in the higher-dimensional scenario.

\paragraph{Changes in covariance (CIC):} For $1\leq i \leq \kappa_1$ and $\kappa_2 +1 \leq i \leq N$, we generate $\bm{x}_i$ from $N\left(\bm{0}, \bm{\Sigma}_1\right).$ For $\kappa_1 +1 \leq i \leq \kappa_2$, we generate $\bm{x}_i$ from $N\left(\bm{0}, \bm{\Sigma}_2\right)$. In the lower-dimensional scenario, we set $p = 4$ and let
$$\bm{\Sigma}_1 = \begin{pmatrix}
1 & 0.8 & 0 & 0\\
0.8 & 1 & 0 & 0\\
0 & 0 & 1 & 0 \\
0 & 0 & 0 & 1\\
\end{pmatrix}, \bm{\Sigma}_2 = \begin{pmatrix}
1 & 0 & 0.8 & 0\\
0 & 1 & 0 & 0\\
0.8 & 0 & 1 & 0 \\
0 & 0 & 0 & 1\\
\end{pmatrix}.$$ 
In the higher-dimensional scenario, we set $p = 8.$ In this case, the diagonal entries of both covariance matrices $\bm{\Sigma}_1$ and $\bm{\Sigma}_2$ are all equal to one. The $(1,2)$ and $(2,1)$ entries of $\bm{\Sigma}_1$ are equal to $0.9,$ but all other off-diagonal entries of $\bm{\Sigma}_1$ are equal to zero. The $(1,3)$ and $(3,1)$ as well as $(2,3)$ and $(3,2)$ entries of $\bm{\Sigma}_2$ are equal to $0.9,$ but all other off-diagonal entries of $\bm{\Sigma}_2$ are equal to zero.

\paragraph{Changes in mean and covariance (CIMC):} For $1\leq i \leq \kappa_1$, we generate $\bm{x}_i$ from  $N\left(\bm{0}, \bm{\Sigma}\right).$ For $\kappa_1 +1 \leq i \leq \kappa_2$, we generate $\bm{x}_i$ from $N\left(\bm{\mu}, \bm{\Sigma}\right).$ 
For $\kappa_2 +1 \leq i \leq N$, we generate $\bm{x}_i$ from $N\left(\bm{\mu}, \bm{I}\right).$ 
In the lower-dimensional scenario, we set $p=4,$ let $\bm{\mu}$ be a $p$-dimensional vector of ones, and define
$$\bm{\Sigma} = \begin{pmatrix}
1 & 0.7 & 0 & 0.7\\
0.7 & 1 & 0 & 0\\
0 & 0 & 1 & 0 \\
0.7 & 0 & 0 & 1\\
\end{pmatrix}.$$ 
In the higher-dimensional scenario, we set $p=8.$ We let the first four entries of $\bm{\mu}$ equal one and the last four entries of $\bm{\mu}$ equal zero. We define $\bm{\Sigma}$ as follows. The diagonal entries of $\bm{\Sigma}$ are all equal to one. The $(1,2)$ and $(2,1)$ as well as $(3,4)$ and $(4,3)$ entries of $\bm{\Sigma}$ are equal to $0.9,$ but all other off-diagonal entries of $\bm{\Sigma}$ are equal to zero. \\

To detect changes in covariance, we transform each observation in the series using a degree-2 polynomial feature embedding $\psi: \mathbb{R}^p \rightarrow \mathbb{R}^{2p+{p\choose2}}$ that maps $\bm{x} = \left(x_1, \dots, x_p\right)$ to
\begin{align}\label{Eq. degree-2 poly-embedding}
\psi(\bm{x}) \coloneqq \left(x_1, \dots, x_p, x_1^2, \dots, x_p^2, x_1x_2, \dots, x_1x_p, x_2x_3, \dots, x_2x_p, \dots, x_{p-1}x_p\right).
\end{align}
That is, instead of applying our method to the series $\bm{x}_1, \ldots, \bm{x}_N,$ we apply it to the series $\psi(\bm{x}_1), \ldots, \psi(\bm{x}_N).$ When $p=4,$ $\psi(\bm{x}_1), \ldots, \psi(\bm{x}_N) \in \mathbb{R}^{14}.$ When $p=8, \psi(\bm{x}_1), \ldots, \psi(\bm{x}_N) \in \mathbb{R}^{44}.$ This transformation step is not necessary for the change in mean scenarios. Each series is centered and standardized prior to being used in the simulation studies to evaluate the performance of each method. In each of the scenarios described above, changes occur in relatively few dimensions compared to the total number of dimensions, motivating our choice to use the horseshoe prior for the regression coefficients in the \texttt{bcmlr} method.

\subsection{Simulation results} \label{Sec. simualtion results}
Tables \ref{Table-unknown-L-paired} and \ref{Table-known-L-paired} present the mean ARIs  (estimated based on 100 simulated data sets) of the different methods in each of the 12 scenarios outlined above. The subscripts $x$ and $\psi$ attached to the mean ARIs indicate the type of input data, with $x$ corresponding to the raw data $\bm{x}_i$ and $\psi$ corresponding to the degree-2 polynomial embedding $\psi(\bm{x}_i)$ given in \eqref{Eq. degree-2 poly-embedding}. The subscripts attached to the column headings indicate the dimension of the input series. (If the polynomial embedding is considered for a particular scenario, then the subscript indicates the dimension of $\psi(\bm{x}_i)$ rather than the dimension of the raw data $\bm{x}_i.$) For each scenario, the highest mean ARI is bolded, while the lowest mean ARI is underlined.

Comparing the mean ARIs when the number of changepoints is unknown (Table \ref{Table-unknown-L-paired}), we observe the following. For the two CIM scenarios, all methods achieve high mean ARIs above 0.95, except \texttt{ClaSP}. For the two CIC scenarios (focusing on the case of polynomial embedded input data), we see that \texttt{kcp} and \texttt{ClaSP} performed poorly (with mean ARIs below 0.1 and 0.6, respectively),  \texttt{MultiRank} and \texttt{changeforest} performed best (with mean ARIs above 0.9), while \texttt{bcmlr} and \texttt{e.divisive} were in the middle. For the two CIMC scenarios (again, focusing on the case of polynomial embedded input data), \texttt{changeforest} performs best by a wide margin (with mean ARIs above 0.95), while \texttt{bcmlr} is second best. 

\begin{table}[htbp]
\centering
\small
\resizebox{\textwidth}{!}{%
\begin{tabular}{lcccccc}
\toprule
& $\textbf{CIM}_{14}$ & $\textbf{CIM}_{40}$
& $\textbf{CIC}_{14}$ & $\textbf{CIC}_{44}$
& $\textbf{CIMC}_{14}$ & $\textbf{CIMC}_{44}$ \\
\midrule
\texttt{bcmlr}
& $0.984_{x}$          & $0.980_{x}$
& $0.804_{\psi}$        & $0.764_{\psi}$
& $0.752_{\psi}$ & $0.630_{\psi}$ \\
\texttt{e.divisive}
& $0.987_{x}$          & $0.967_{x}$
& $0.055_x, 0.886_{\psi}$        & $0.022_x, 0.704_{\psi}$
& $0.538_x, 0.535_{\psi}$        & $0.532_x, 0.532_{\psi}$ \\
\texttt{kcp}
& $\mathbf{1.000}_{x}$  & $\mathbf{1.000}_{x}$
& $0.088_x, 0.095_{\psi}$        & $0.010_x, 0.031_{\psi}$
& $0.626_x, 0.609_{\psi}$        & $0.555_x, 0.560_{\psi}$ \\
\texttt{MultiRank}
& $0.989_{x}$           & $0.989_{x}$
& $\underline{0.005}_x, \mathbf{0.954}_{\psi}$ & $\underline{0.004}_x, 0.931_{\psi}$
& $0.536_x, 0.741_{\psi}$        & $0.546_x, 0.611_{\psi}$ \\
\texttt{ClaSP}
& $\underline{0.263}_{x}$           & $\underline{0.287}_{x}$
& $0.598_x, 0.565_{\psi}$ & $0.526_x, 0.497_{\psi}$
& $\underline{0.430}_x, 0.497_{\psi}$        & $\underline{0.399}_x, 0.471_{\psi}$ \\
\texttt{changeforest}
& $0.990_{x}$           & $0.984_{x}$
& $0.873_x, 0.952_{\psi}$ & $0.574_x, \mathbf{0.953}_{\psi}$
& $\mathbf{0.974}_x, 0.970_{\psi}$        & $0.966_x, \mathbf{0.967}_{\psi}$ \\
\bottomrule
\end{tabular}
}
\caption{Mean ARIs when the number of changepoints is unknown.}
\label{Table-unknown-L-paired}
\end{table}

Looking at the mean ARIs when the number of changepoints is known (Table~\ref{Table-known-L-paired}), we immediately notice that the mean ARIs are higher than they were in the case when the number of changepoints was unknown, as we should expect. For the two CIM scenarios, all methods achieve high mean ARIs above 0.98, except \texttt{ClaSP}. For the two CIC scenarios (with polynomial embedded input data), \texttt{bcmlr} and \texttt{MultiRank} perform best (with mean ARIs above 0.97). For the two CIMC scenarios (with polynomial embedded input data), \texttt{MultiRank} performs best with \texttt{bcmlr} a close second. The \texttt{changeforest} method is excluded from the comparison presented in Table~\ref{Table-known-L-paired} because the available implementation does not allow the user to specify a known number of changepoints. 

\begin{table}[htbp]
\centering
\small
\resizebox{\textwidth}{!}{%
\begin{tabular}{lcccccc}
\toprule
& $\textbf{CIM}_{14}$ & $\textbf{CIM}_{40}$
& $\textbf{CIC}_{14}$ & $\textbf{CIC}_{44}$
& $\textbf{CIMC}_{14}$ & $\textbf{CIMC}_{44}$ \\
\midrule
\texttt{bcmlr}
& $0.998_{x}$           & $0.998_{x}$
& $0.974_{\psi}$        & $\mathbf{0.981}_{\psi}$
& $0.944_{\psi}$        & $0.954_{\psi}$ \\
\texttt{e.divisive}
& $0.989_{x}$           & $0.989_{x}$
& $0.400_x, 0.927_{\psi}$        & $0.325_x, 0.878_{\psi}$
& $0.565_x, 0.621_{\psi}$        & $0.484_x, 0.647_{\psi}$ \\
\texttt{kcp}
& $\mathbf{1.000}_{x}$  & $\mathbf{1.000}_{x}$
& $0.876_x, 0.954_{\psi}$        & $0.628_x, 0.843_{\psi}$
& $0.906_x, 0.881_{\psi}$        & $0.821_x, 0.896_{\psi}$ \\
\texttt{MultiRank}
& $\mathbf{1.000}_{x}$  & $\mathbf{1.000}_{x}$
& $\underline{0.315}_x, \mathbf{0.979}_{\psi}$ & $\underline{0.309}_x, \mathbf{0.981}_{\psi}$
& $0.679_x, \mathbf{0.970}_{\psi}$ & $0.708_x, \mathbf{0.976}_{\psi}$ \\
\texttt{ClaSP}
& $\underline{0.339}_{x}$           & $\underline{0.347}_{x}$
& $0.806_x, 0.777_{\psi}$ & $0.670_x, 0.652_{\psi}$
& $\underline{0.482}_x, 0.558_{\psi}$        & $\underline{0.461}_x, 0.510_{\psi}$ \\
\bottomrule
\end{tabular}
}
\caption{Mean ARIs when the number of changepoints is known.}
\label{Table-known-L-paired}
\end{table}

In terms of mean ARIs, \texttt{bcmlr} is competitive, but not always the top performer. This performance is not surprising given that the \texttt{bcmlr} loss function is constructed from a parametric multinomial logistic regression, whereas many of the competing methods are nonparametric or leverage black box machine learning classifiers. However, \texttt{bcmlr} provides rich information that goes beyond point estimates of changepoint locations. The choice of method should depend on the user’s objective. If maximizing changepoint estimation accuracy is the primary goal, Table \ref{Table-unknown-L-paired} suggests that \texttt{changeforest} may be the best choice. If one wants changepoint estimation accuracy as well as uncertainty quantification and insight into the changes that may have occured, \texttt{bcmlr} may be preferable.

    \section{Real data applications}\label{Sec. real data app}

In this section, we demonstrate via real data applications that the \texttt{bcmlr} method is well-suited for analyzing subtle changes in complex data. In 
the Supplementary Material, we also apply \texttt{bcmlr} to annotated data series from \citet{van2020evaluation}. These examples serve as a middle ground between simulation studies with known ground truth and our main applications in Sections \ref{Sec. DJIA} and \ref{Sec. nanoparticle video}.

    \subsection{DJIA negative correlation networks}
\label{Sec. DJIA}




\begin{figure}[h]
    \centering
    \includegraphics[width=\textwidth]{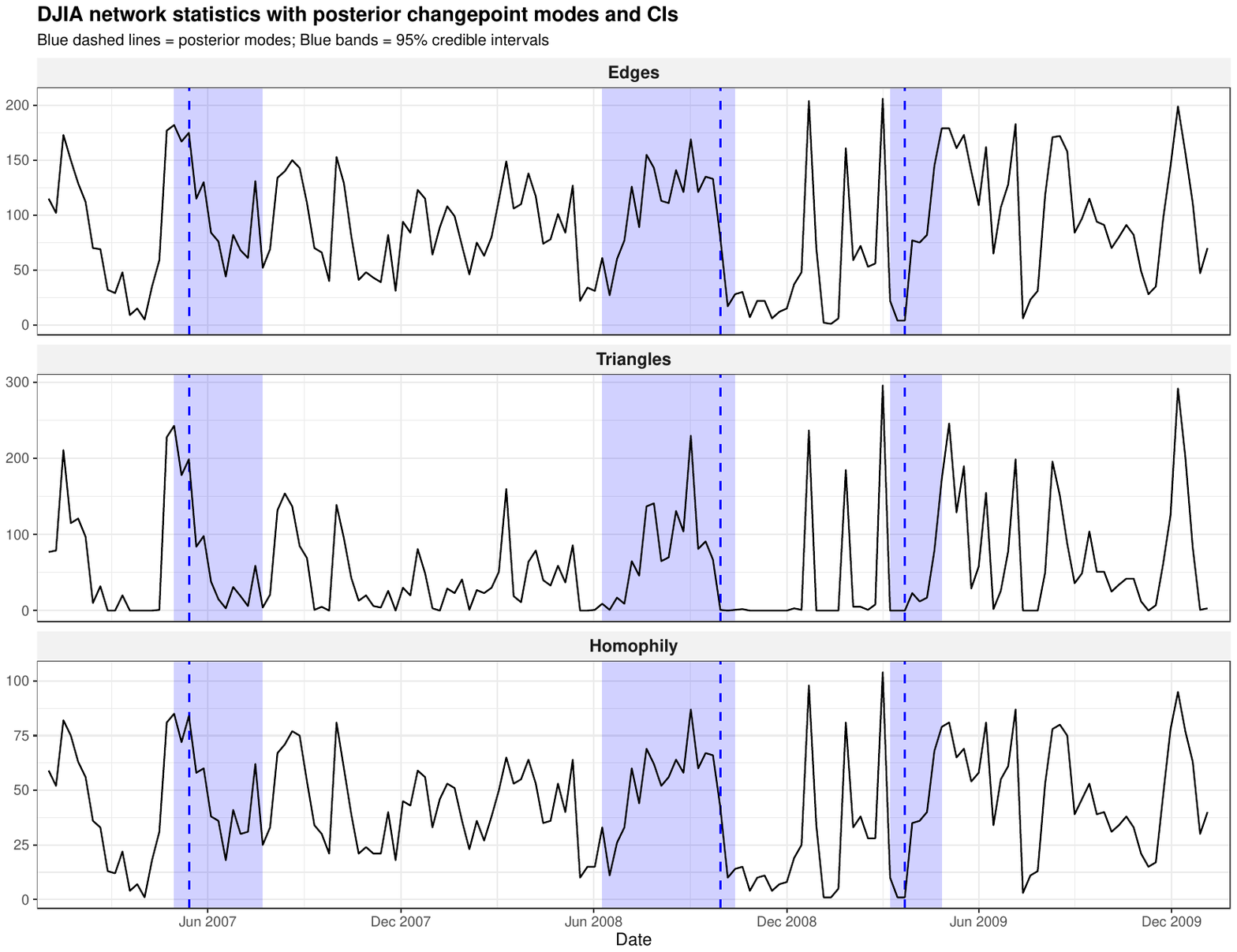}
    \caption{The DJIA network statistics with posterior mode estimates of changepoints (blue dashed lines) along with 95\% credible intervals (blue bands) obtained from applying \texttt{bcmlr}. }
    \label{fig:DJIA-network-time-series}
\end{figure}

Correlations between stock returns are related to market trends \citep{longin2001extreme}. In particular, correlations increase in bear markets such as the 2008 financial crisis. \citet{kei2023change} conducted changepoint analysis based on network data derived from correlations among stock returns during the 2008 financial crisis. In this section, we re-examine this data with our proposed method to illustrate the rich information it provides for changepoint analysis and to compare the changepoints found by our method with those found by \citet{kei2023change} and the methods considered in Section~\ref{Sec. Simulation}.  

To construct adjacency matrices representing network data, \citet{kei2023change} considered weekly log returns of 29 stocks in the Dow Jones Industrial Average (DJIA) from 2007-01-01 to 2010-01-04, a time frame covering the 2008 financial crisis. The data are available in the \texttt{ecp} package \citep{james2015ecp}. The authors first computed sample correlation matrices of the log returns for each four-week rolling window. The adjacency matrices were constructed from these sample correlation matrices by setting entries equal to one (indicating an edge) if the corresponding sample correlation was negative and setting entries equal to zero (indicating no edge) otherwise. This process yielded a series of $N=158$ adjacency matrices of dimension $29 \times 29.$

For each (undirected) adjacency matrix, \citet{kei2023change} computed three network statistics—the number of edges, the number of triangles, and homophily for risk orientation (see Figure~\ref{fig:DJIA-network-time-series}). To compute homophily for risk orientation, they first assigned labels to each node (stock) based on its cumulative node degree. Stocks with cumulative degrees above the median cumulative degree were labeled ``Hedging-Prone", while the remaining stocks were labeled ``Market-Following". ``Hedging-Prone" stocks tend to play a defensive role relative to market dynamics, while ``Market-Following" stocks align more closely with broader market movements. Homophily for risk orientation is defined as the number of edges whose endpoints share the same label. This statistic captures the tendency for stocks with the same risk orientation to be negatively correlated with each other. \citet{kei2023change} used these three network statistics as input data for their method, \texttt{CPDstergm}. To allow for comparison, we use the same network statistics as input data for our proposed method and the competing methods from the simulation studies in Section~\ref{Sec. Simulation}. 

\begin{figure}[h]
    \centering
    \includegraphics[width=\textwidth]{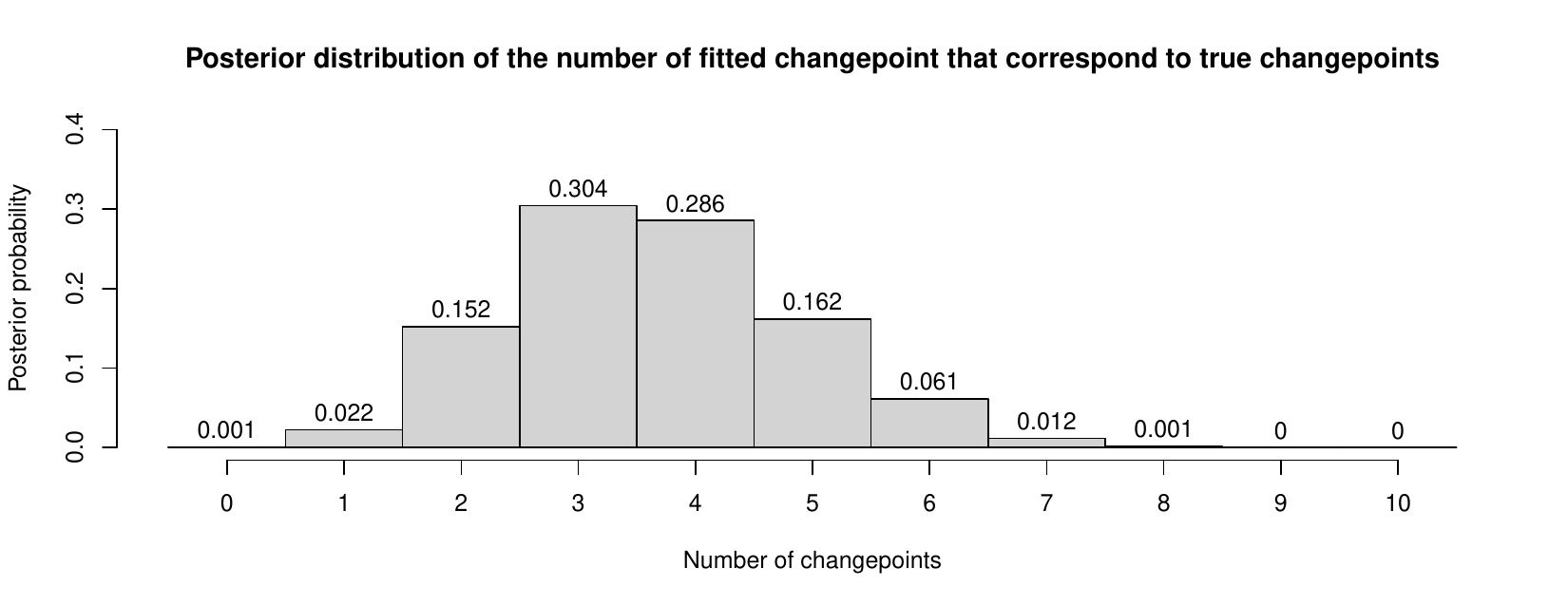}
    \caption{The posterior distribution of the number of 10 fitted changepoints that correspond to true changepoints based on 5000 posterior samples (after 10000 burn-in samples) obtained from applying the approach in Section \ref{Sec. select num of CPs} to select the number of changepoints on the DJIA series.}
    \label{fig:DJIA-numCPdistribution}
\end{figure}
Since the number of changepoints $L_{\text{true}}$ in the network series is unknown, we apply the approach in Section \ref{Sec. select num of CPs} to select the number of changepoints. We set $L_{\text{fitted}} = 10$ (this is the upper bound for $L_{\text{true}}$), the confidence level $1-\alpha = 0.9$, the AUC threshold $\tau = 0.5$, the minimum segment length $m = 10$, and we hold out every 5th observation ($\zeta = 5$) for AUC calculation. Since the data is low-dimensional, we use a multivariate Gaussian $N\left(\bm{0}, 3\bm{I}\right)$ as the prior distribution for the regression coefficients. We center and standardize the network statistics before applying our method. We initialize the changepoints at evenly spaced locations along the series and run the Gibbs sampler for $T = 15{,}000$ iterations, with a burn-in period of $T_0 = 10{,}000$ iterations. Figure~\ref{fig:DJIA-numCPdistribution} shows a histogram approximating the posterior distribution of the number of fitted changepoints that correspond to true changepoints, in the sense discussed in Section~\ref{Sec. select num of CPs}. From the histogram, we see that the posterior distribution does not concentrate on one value, indicating uncertainty about the number of fitted changepoints that correspond to true changepoints. (In light of this finding, it is not surprising that some of the other methods identify different numbers of changepoints, as shown in Table~\ref{Table: financial events}.) In scenarios like this, one may want to refit the \texttt{bcmlr} method for each of the most probable values of $L.$ In this section, however, we take the posterior mode $\hat{L}_{\text{true}} = 3$ as our point estimate of $L_{\text{true}}$ and condition all subsequent analysis on the number of changepoints being equal to three. 
\begin{table}[h]
\centering
\small
\begin{tabular}{c|cccc}
\hline
Real event dates & \texttt{bcmlr} & \texttt{CPDstergm} & \texttt{e.divisive} & \texttt{MultiRank} \\
\hline
(i)~~2007-04-02 & 2007-05-14 & 2007-04-23 & -- & -- \\
(ii)~2008-09-15 & 2008-09-29 & 2008-10-06 & 2008-10-06 & 2008-09-15 \\
(iii)~2009-03-09 & 2009-03-23 & 2009-04-20 & 2009-04-20 & 2009-04-06 \\
-- & -- & -- & -- & 2009-05-18 \\
-- & -- & -- & -- & 2009-06-29 \\
-- & -- & -- & -- & 2009-11-16 \\
-- & -- & -- & -- & 2009-12-28 \\
\hline
\end{tabular}
\caption{Estimated changepoints from each method that identified at least one changepoint compared with the dates of real events reviewed by \citet{kei2023change}.}
\label{Table: financial events}
\end{table}

We now compare the changepoints detected by the \texttt{CPDstergm} method of \citet{kei2023change} to those detected by our proposed method. \citet{kei2023change} found three changepoints corresponding to the following dates: 2007-04-23, 2008-10-06, and 2009-04-20. The authors linked these changepoints to three approximately coincident events: (i) New Century Financial Corporation filed for bankruptcy on 2007-04-02, (ii) Lehman Brothers filed for bankruptcy on 2008-09-15, and (iii) the DJIA index bottomed out on 2009-03-09. The \texttt{bcmlr} method, applied to the entire series with $L = 3$, identifies three changepoints corresponding to the following dates: 2007-05-14, 2008-09-29, and 2009-03-23. Compared to the results from \texttt{CPDstergm}, the latter two changepoints detected by \texttt{bcmlr} align more closely with the dates of events (ii) and (iii). Figure~\ref{fig:DJIA-network-time-series} displays the posterior mode estimates of the changepoints (indicated by blue vertical dashed lines) along with 95\% credible intervals (indicated by blue bands). Table~\ref{Table: financial events} shows the estimated changepoints from each method that identified at least one changepoint compared to the dates of events (i)-(iii).

\begin{figure}[h]
  \centering
  \includegraphics[width=\linewidth]{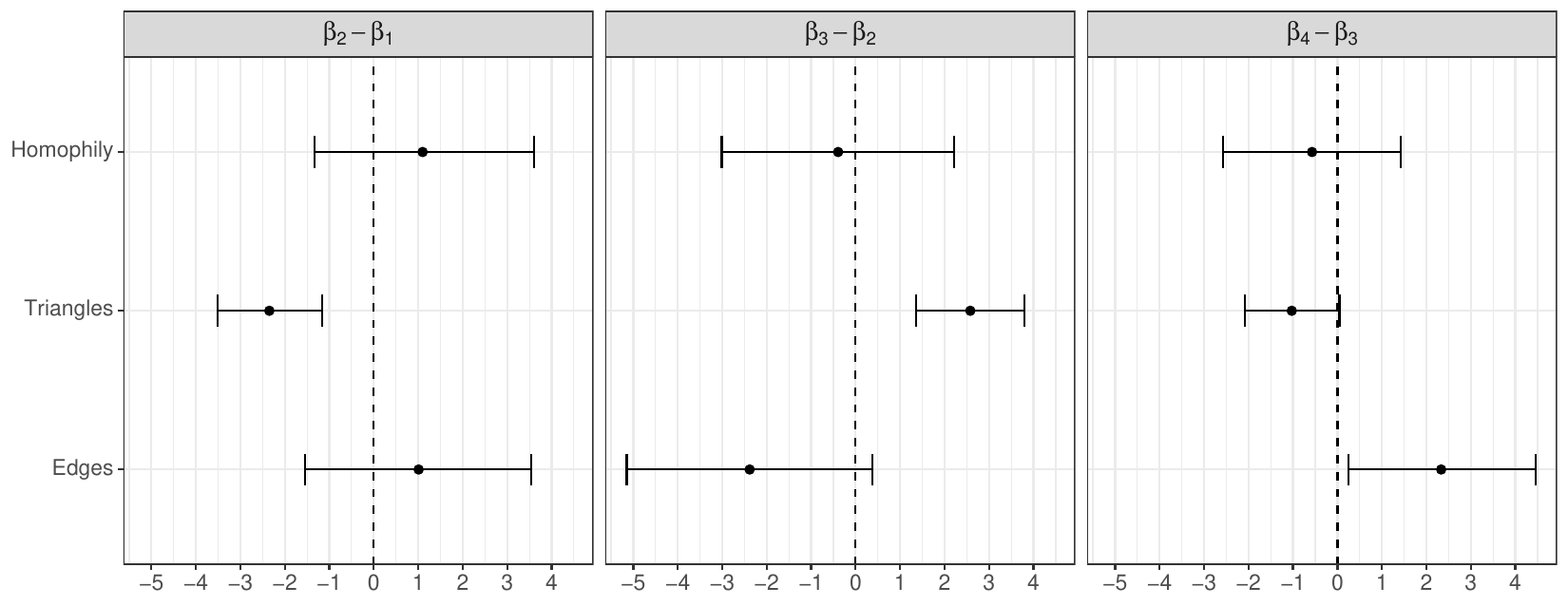}
  \caption{Each plot shows 95\% credible intervals for difference
  $\bm{\beta}_{l+1}-\bm{\beta}_{l}$ for $l \in \{1, \dots, 3\}$. The three dimensions of each difference vector relate to edges, triangles, and homophily for risk orientation. The black dots represent posterior means.}
  \label{fig:DJIA-beta-diff}
\end{figure}

\begin{figure}[h]
  \centering
  \includegraphics[width=\linewidth]{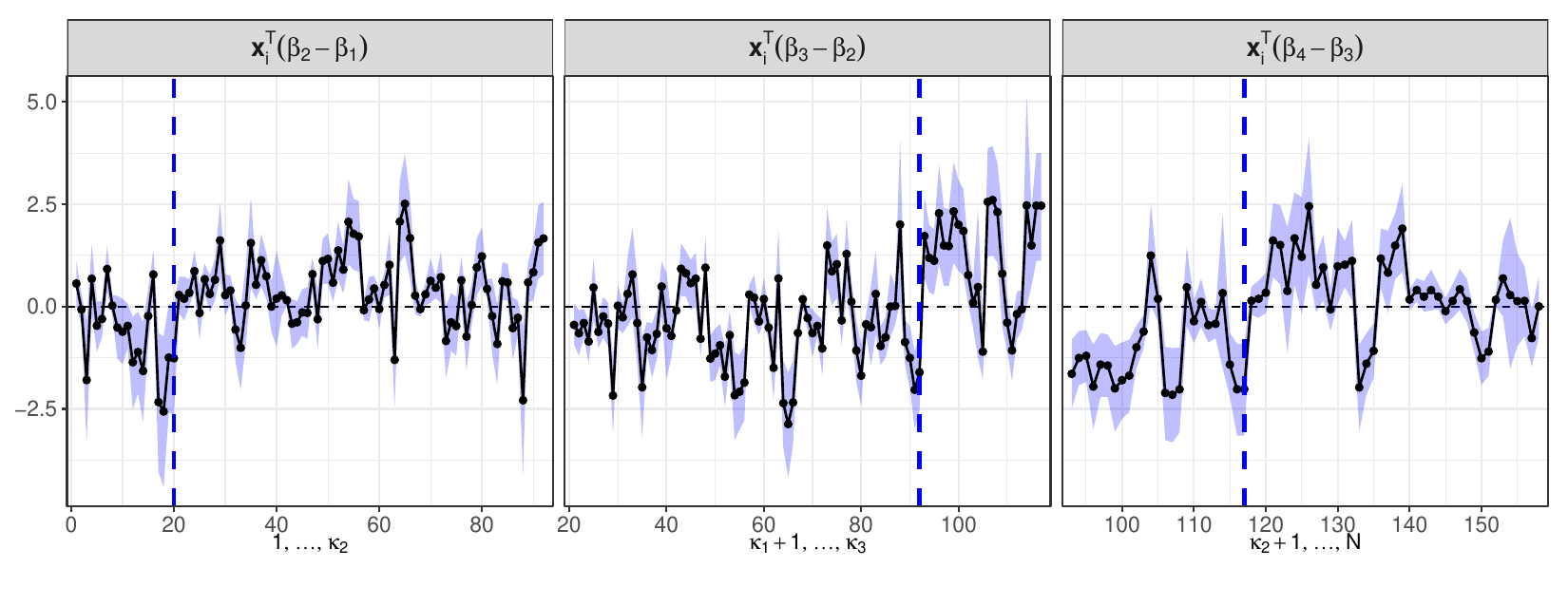}
  \caption{Each plot shows posterior means of $\bm{x}_i^{\top}\left(\bm{\beta}_{l+1}-\bm{\beta}_l\right)$, along with 95\% pointwise credible intervals (blue bands), based on the data in neighboring segments on either side
  of the changepoint $\kappa_l$. The blue dashed lines represent the posterior mode estimates of the changepoints.}
  \label{fig:DJIA-log-prob-diff}
\end{figure}

We can interpret which covariates contributed to each change by examining posterior summaries of the regression coefficients, in much the same way that we would interpret the coefficients of a multinomial logistic regression. The difference $\bm{\beta}_{l+1} -\bm{\beta}_{l}$ is particularly relevant for interpreting the $l$th changepoint. Figure~\ref{fig:DJIA-beta-diff} presents point estimates and 95\% credible intervals of these differences for each $l \in \{1, \dots, 3\}$. As an illustration of how we can interpret these differences, consider the third subplot in Figure~\ref{fig:DJIA-beta-diff}, corresponding to the difference $\bm{\beta}_{4} -\bm{\beta}_3.$ We see that the lower bound of the credible interval for the number of edges lies above zero, whereas the credible intervals for the number of triangles and homophily for risk orientation both include zero. This suggests that the third changepoint corresponds to an increase in the number of edges. An increase in the number of negatively correlated log returns suggests reduced market risk, which aligns with our knowledge that this changepoint occurred during the period in which the DJIA index bottomed out and began to recover after a 17-month bear market. One can interpret the other changepoints in a similar fashion based on their respective subplots. It is also possible to interpret differences between non-adjacent segments by examining the differences between non-consecutive coefficient vectors. For example,  the difference $\bm{\beta}_4 -\bm{\beta}_1$ contains information about how the the fourth segment differs from the first segment.

In the multinomial logistic regression model implied by our loss function, the conditional probability that $\bm{x}_i$ belongs to class $l+1$ (i.e. the segment immediately following $\kappa_l$) rather than class $l$ (i.e. the segment immediately preceding $\kappa_l$), given that it belongs to one of these two classes, depends on the data through the inner product $\bm{x}_i^\top (\bm{\beta}_{l+1} - \bm{\beta}_{l}).$ Figure~\ref{fig:DJIA-log-prob-diff} plots the posterior mean and pointwise 95\% credible intervals of this quantity for each changepoint. These one-dimensional summaries make it easier to visualize the changes occurring at each changepoint. 

As we alluded to earlier, the competing methods from the simulation studies in Section~\ref{Sec. Simulation} identify a variety of different collections of changepoints when applied to the network statistics. The \texttt{e.divisive} method (with the minimum segment length set to 10) detects two changepoints that coincide with the latter two changepoint estimates of \texttt{CPDstergm}. The three methods \texttt{kcp} (with the maximum number of changepoints set to 10), \texttt{ClaSP} (with a validation classification score threshold of 0.75), and \texttt{changeforest} detect no changepoints. In contrast, \texttt{MultiRank} (with the maximum number of changepoints set to 10) detects six changepoints, with the first two occurring around events (i) and (ii).

    \subsection{Nanoparticle video} \label{Sec. nanoparticle video}

In this section, we apply our proposed method to the denoised nanoparticle video appearing in Figure 4 of \citet{crozier2025}. The authors of that article used topological data analysis to heuristically characterize nanoparticles into ordered and disordered/unstable states. The various behaviors of the nanoparticles in ordered states were further characterized by their angular and facial structure. However, all of this analysis was done on a frame-by-frame basis by experts. We will go a step further by using our proposed method to identify changes in state and characterize these changes in a more automated, algorithmic fashion via the estimates of the regression coefficients $\bm{\beta}_1, \dots, \bm{\beta}_J.$  

A distinct advantage of the multinomial logistic approach that we pursue here (as opposed to the binomial logistic approach applied to consecutive segments, as pursued in \citealp{thomas2024bayesian}) is that it allows us to compare the properties of a given segment to those of the other segments. This enables us to better understand the behavior of our data series within each homogeneous segment. As a heuristic method of characterizing the behavior in each segment, we apply $k$-means clustering to the posterior mean estimates of $\bm{\beta}_1, \dots, \bm{\beta}_J$, selecting the smallest $k$ which produces alternating ``regimes'' (i.e. sequences of frames with similar behavior). Note that, as an extension of our generalized Bayesian method, we could specify a certain number of regimes and associate various segments to them in an unsupervised fashion---akin to a hidden Markov model---but we leave this approach to future work. 

Before moving on, we enumerate the parameters used when applying our method. The number of changepoints is unknown, so we set $L_\text{fitted} = 12$. Furthermore, we set the confidence level to $1-\alpha = 0.9$, the AUC threshold to $\tau = 0.75$, the minimum segment length to $m = 30,$ and held out every 5th observation $(\zeta = 5)$ for AUC computation. Furthermore, we ran the Gibbs sampler for $T = 5,000$ iterations with a burn-in period of $T_0 = 2,500$. The posterior mode of the number of changepoints was $\hat{L}_{\text{true}} = 8$, which had a posterior probability of $0.649$. We employed this value as our estimate of $L_{\text{true}}$ in this case. 

\begin{figure}[t]
    \centering
    \includegraphics[width=\textwidth]{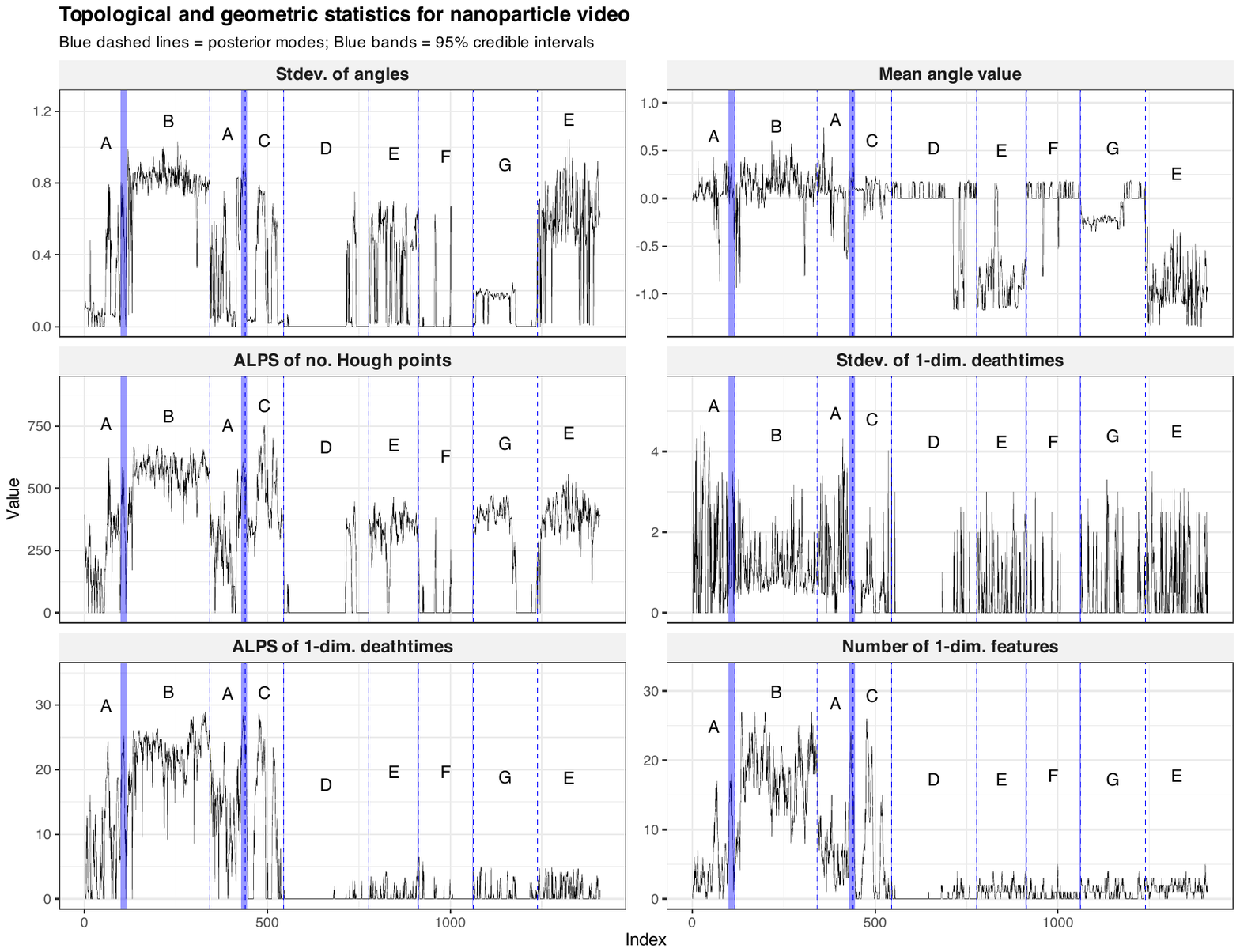}
    \caption{Plot of 6 of the 7 features used to detect nanoparticle activity and tilt in the denoised video from Figure 4 of \cite{crozier2025}. Blue dashed lines are posterior modes and blue bands are 95\% credible intervals.}
    \label{fig:NanoPlot}
\end{figure}

\begin{figure}[t]
    \centering
    \includegraphics[width=\textwidth]{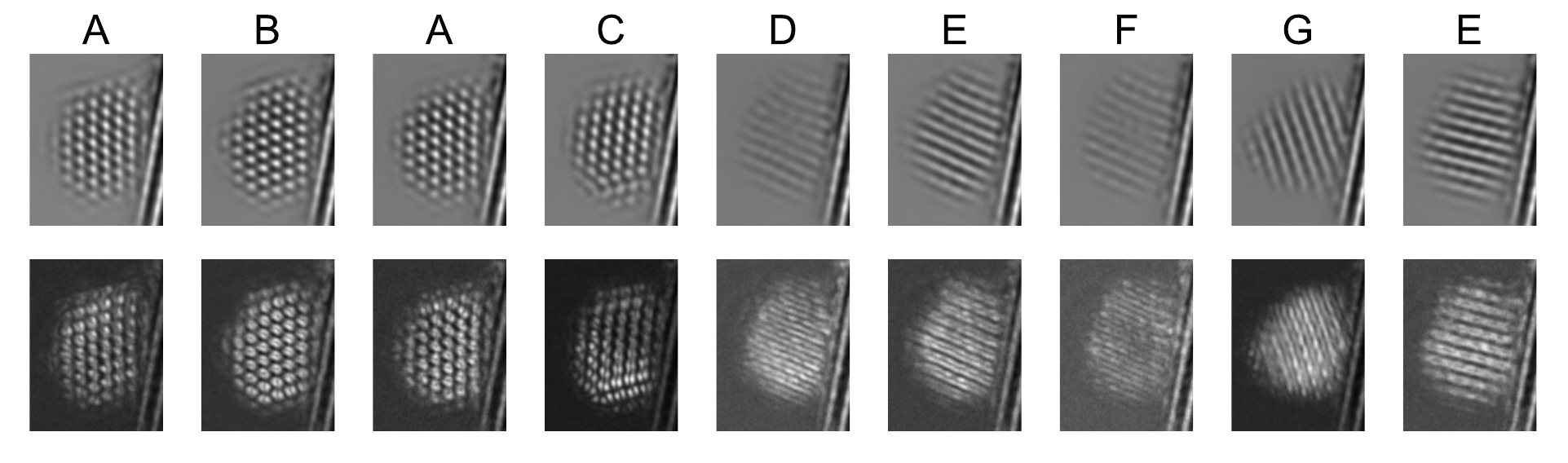}
    \caption{(First row) Pixelwise means for each regime identified by \texttt{bcmlr}. (Second row) Pixelwise standard deviations for each regime identified by \texttt{bcmlr}. We can see these correspond well to what is seen in Figure~\ref{fig:NanoPlot}. All of these images are normalized to have mean 1 for ease of comparison.}
    \label{fig:NanoRegime}
\end{figure}

To conduct our analysis of the denoised nanoparticle video dynamics, we chose a combined topological/geometric feature embedding of statistics. Before analyzing the video, we smoothed each frame in the video according to a Gaussian filter with $\sigma=2$, binarized it via cross-entropy thresholding \citep{li1993}, and then removed small 8-connected black regions in the binary image (of area less than 0.06\% of the image size) to reduce noise in our statistics. At this point, we summarized our image geometrically by using statistics of the Hough line transform \citep{hough_transform}, and topologically by applying the signed distance filtration to each preprocessed, binarized image \citep{garin2019}. Specifically, for a white pixel, the signed (Manhattan) distance filtration assigns a pixel intensity equal to the minimum $L_1$ distance (minus 1) to a black pixel; for a black pixel, it assigns a pixel intensity equal to the minimum $L_1$ distance to any white pixel. We calculated 4 geometric statistics from the Hough transform and 3 topological statistics. We chose independent $N(\bm{0}, 3\mathbf{I})$ priors for each of the regression coefficient vectors.

For the Hough transform statistics, we identified the 8-connected components in the binarized image and applied the Hough transform to each of these individually, thresholding the accumulator in Hough space using a value of 150. Using the aggregated peaks of the Hough line transform from all the connected components, we then calculated the standard deviation of the angle values of the peaks (`Stdev. of angles' in Figure~\ref{fig:NanoPlot}); the mean angle of resulting peaks (`Mean angle value' in Figure~\ref{fig:NanoPlot}); the ALPS statistic \citep{thomas2023} of the values in the aggregated accumulator array (i.e. `ALPS of no. Hough points' in Figure~\ref{fig:NanoPlot}); and the total number of Hough points (not depicted in Figure~\ref{fig:NanoPlot}). These values capture a specific crystallographic structure in the nanoparticle bulk which appears in the images as parallel lines (so-called \emph{Miller planes}); it also describes the angles these ``lines'' take relative to the image origin (see right five plots in Figure~\ref{fig:NanoRegime}).

As previously mentioned, we assessed dynamics with topological statistics as well. To get a finer-grained picture of how the nanoparticle evolves, we use the signed (Manhattan) distance filtration \citep{garin2019}. This  gives a better idea of the visual size of topological features than the sublevel set filtration employed in \citet{thomas2023} and \citet{crozier2025}. We aim to describe the lattice structure via the bright spots of the nanoparticle images (cf. top row of Figure~\ref{fig:NanoRegime}), and we do so using white regions in the subsequently binarized image that are surrounded by black pixels in the binary image. That is, we use ``quadrant IV'' features (as detailed in \citealp{moon2023}) and investigate their death times---which captures the $L_1$ radii of the white dots in regimes \textsf{A, B, C} or the $L_1$ width of the lines\footnote{For example, this captures the ``lattice plane separation distance'' mentioned in \cite{vincent2021}.} in regimes \textsf{D, E, F,} or \textsf{G}. We calculated the standard deviation of the 1-dimensional deathtimes (`Stdev. of 1-dim. deathtimes in Figure~\ref{fig:NanoPlot}); the ALPS statistic of 1-dim. deathtimes (`ALPS of 1-dim. deathtimes in Figure~\ref{fig:NanoPlot}); and the number of 1-dimensional features in quadrant IV (`Number of 1-dim. features' in Figure~\ref{fig:NanoPlot}). Heuristically, these tracked the regularity of the lattice structure and the number of ``points'' in the lattice (with ALPS being a continuous measure of the `Number of 1-dim. features'), respectively. 

In Figure~\ref{fig:NanoRegime}, we can see that regime \textsf{A} is characterized by a regular hexagonal lattice structure and lower fluxionality (particle movement) compared to regime \textsf{B}, as the second row of Figure~\ref{fig:NanoRegime} for regime \textsf{B} is brighter (indicating greater pixelwise variance) than that of regime A to the left and right (though the size of the atomic features seems more regular in regime \textsf{B} than regime \textsf{A} by looking at the `Stdev. of 1-dim deathimes'). Regime \textsf{C} seems to display a greater degree of activity at the lower surface of the nanoparticle (i.e. the second row is brighter as we go lower). Regimes \textsf{D} and \textsf{F} are both highly fluxional (with nanoparticle structure rarely being seen in the video due to rapid movement) and `Mean angle value' indicates that perhaps regime \textsf{D} is a little more structured than regime \textsf{F}. Regimes \textsf{E} and \textsf{G} are stable but the particle is oriented at different angles. All of these qualitative nanoparticle configurations are captured via our summaries, and in an interpretable manner (as opposed to a Fourier transform).

    \section{Discussion} \label{Sec. Discussion}
We presented a generalized Bayesian multiple changepoint method with a loss function inspired by multinomial logistic regression. The method does not require specification of the data-generating process and avoids overly restrictive assumptions on the nature of the changepoints. From the joint posterior distribution, we can make simultaneous inference on the locations of changepoints and the coefficients of a multinomial logistic regression for distinguishing data across homogeneous segments. 

To select the number of changepoints, we introduced an approach based on posterior summaries that leverages the multinomial logistic regression model implied by the loss function to assess whether data on either side of a potential changepoint can be reliably distinguished. To simulate from the generalized posterior distribution, we constructed a Gibbs sampler via Pólya-Gamma data augmentation. We assessed the accuracy of our method through simulation studies with different types of changes. 

We also applied the method to network data derived from DJIA stock returns and a nanoparticle video. In the nanoparticle video example, we saw that the proposed changepoint method, along with a novel geometric and topological embedding, provides an effective means of 
characterizing atomic-level fluxional behavior of nanoparticles in transmission electron microscopy. In the network data example, we saw that the proposed method offers uncertainty quantification and allows for straightforward interpretation of changepoints. As future work, we intend to pursue an extension that can accommodate adjacency matrices directly and incorporate appropriate dimension reduction rather than relying upon well-chosen network summary statistics.

\section*{Acknowledgements} %
The last author was supported by the National Science Foundation (DMS-2515376).

    \appendix
\numberwithin{equation}{section}
\numberwithin{table}{section}

\bibliographystyle{apalike}
\bibliography{Other/Refs}

@article{thomas2024bayesian,
  author  = {Thomas, Andrew M. and Jauch, Michael and Matteson, David S.},
  title   = {Bayesian Changepoint Detection via Logistic Regression and the Topological Analysis of Image Series},
  journal = {Technometrics},
  year    = {2025},
  volume  = {67},
  number  = {4},
  pages   = {639--705},
  doi     = {10.1080/00401706.2025.2515928}
}

@article{bhattacharya2016fast,
  author  = {Bhattacharya, Anirban and Chakraborty, Antik and Mallick, Bani K.},
  title   = {Fast sampling with {Gaussian} scale-mixture priors in high-dimensional regression},
  journal = {Biometrika},
  year    = {2016},
  volume  = {103},
  number  = {4},
  pages   = {985--991},
  doi     = {10.1093/biomet/asw042}
}

@article{bellman1954theory,
  author  = {Bellman, Richard},
  title   = {The theory of dynamic programming},
  journal = {Bulletin of the American Mathematical Society},
  year    = {1954},
  volume  = {60},
  number  = {6},
  pages   = {503--515},
  doi     = {10.1090/S0002-9904-1954-09848-8}
}

@article{page1954continuous,
  author  = {Page, Ewan S.},
  title   = {Continuous inspection schemes},
  journal = {Biometrika},
  year    = {1954},
  volume  = {41},
  number  = {1/2},
  pages   = {100--115}
}

@article{kanrar2025model,
  title={Model-free Change-point Detection using {AUC} of a Classifier},
  author={Kanrar, Rohit and Jiang, Feiyu and Cai, Zhanrui},
  journal={Journal of Machine Learning Research},
  volume={26},
  number={190},
  pages={1--50},
  year={2025}
}

@article{van2020evaluation,
  author  = {Van den Burg, Gerrit J. J. and Williams, Christopher K. I.},
  title   = {An Evaluation of Change Point Detection Algorithms},
  journal = {arXiv preprint arXiv:2003.06222},
  year    = {2020}
}

@article{matteson2014nonparametric,
  title={A nonparametric approach for multiple change point analysis of multivariate data},
  author={Matteson, David S and James, Nicholas A},
  journal={Journal of the American Statistical Association},
  volume={109},
  number={505},
  pages={334--345},
  year={2014},
  publisher={Taylor \& Francis}
}

@article{vostrikova1981detecting,
  author  = {Vostrikova, L. Yu.},
  title   = {Detecting ``disorder'' in multidimensional random processes},
  journal = {Doklady Akademii Nauk SSSR},
  year    = {1981},
  volume  = {259},
  number  = {2},
  pages   = {270--274}
}

@article{szekely2005hierarchical,
  author  = {Sz{\'e}kely, G{\'a}bor J. and Rizzo, Maria L.},
  title   = {Hierarchical clustering via joint between-within distances: Extending {Ward's} minimum variance method},
  journal = {Journal of Classification},
  year    = {2005},
  volume  = {22},
  number  = {2},
  pages   = {151--184},
  doi     = {10.1007/s00357-005-0012-9}
}

@article{james2015ecp,
  title={ecp: An {R} package for nonparametric multiple change point analysis of multivariate data},
  author={James, Nicholas A and Matteson, David S},
  journal={Journal of Statistical Software},
  volume={62},
  pages={1--25},
  year={2015}
}

@article{arlot2019kernel,
  title={A kernel multiple change-point algorithm via model selection},
  author={Arlot, Sylvain and Celisse, Alain and Harchaoui, Zaid},
  journal={Journal of Machine Learning Research},
  volume={20},
  number={162},
  pages={1--56},
  year={2019}
}

@article{lung2015homogeneity,
  author  = {Lung-Yut-Fong, Alexandre and L{\'e}vy-Leduc, C{\'e}line and Capp{\'e}, Olivier},
  title   = {Homogeneity and change-point detection tests for multivariate data using rank statistics},
  journal = {Journal de la Soci{\'e}t{\'e} Fran{\c{c}}aise de Statistique},
  year    = {2015},
  volume  = {156},
  number  = {4},
  pages   = {133--162},
  doi     = {10.24138/jss.v156i4.216}
}

@article{londschien2023random,
  title={Random forests for change point detection},
  author={Londschien, Malte and B{\"u}hlmann, Peter and Kov{\'a}cs, Solt},
  journal={Journal of Machine Learning Research},
  volume={24},
  number={216},
  pages={1--45},
  year={2023}
}

@article{hocking2013learning,
  author  = {Hocking, Toby Dylan and Schleiermacher, Gudrun and Janoueix-Lerosey, Isabelle and Boeva, Valentina and Cappo, Julie and Delattre, Olivier and Bach, Francis and Vert, Jean-Philippe},
  title   = {Learning smoothing models of copy number profiles using breakpoint annotations},
  journal = {BMC Bioinformatics},
  year    = {2013},
  volume  = {14},
  pages   = {164},
  doi     = {10.1186/1471-2105-14-164}
}

@article{verbesselt2010detecting,
  author  = {Verbesselt, Jan and Hyndman, Rob J. and Newnham, Glenn and Culvenor, Darius},
  title   = {Detecting trend and seasonal changes in satellite image time series},
  journal = {Remote Sensing of Environment},
  year    = {2010},
  volume  = {114},
  number  = {1},
  pages   = {106--115},
  doi     = {10.1016/j.rse.2009.08.014}
}

@article{banerjee2020change,
  author  = {Banerjee, Sayantan and Guhathakurta, Kousik},
  title   = {Change-point analysis in financial networks},
  journal = {Stat},
  year    = {2020},
  volume  = {9},
  number  = {1},
  pages   = {e269},
  doi     = {10.1002/sta4.269}
}

@article{ermshaus2023clasp,
  title={{ClaSP}: parameter-free time series segmentation},
  author={Ermshaus, Arik and Sch{\"a}fer, Patrick and Leser, Ulf},
  journal={Data Mining and Knowledge Discovery},
  volume={37},
  number={3},
  pages={1262--1300},
  year={2023},
  publisher={Springer}
}

@article{azzalini2014clustering,
  title={Clustering via nonparametric density estimation: The {R} package {pdfCluster}},
  author={Azzalini, Adelchi and Menardi, Giovanna},
  journal={Journal of Statistical Software},
  volume={57},
  pages={1--26},
  year={2014}
}

@article{pedregosa2011scikit,
  author  = {Pedregosa, Fabian and Varoquaux, Ga{\"e}l and Gramfort, Alexandre and Michel, Vincent and Thirion, Bertrand and Grisel, Olivier and Blondel, Mathieu and Prettenhofer, Peter and Weiss, Ron and Dubourg, Vincent and others},
  title   = {{Scikit-learn}: {Machine Learning} in {Python}},
  journal = {Journal of Machine Learning Research},
  year    = {2011},
  volume  = {12},
  pages   = {2825--2830}
}

@article{morey1984measurement,
  author  = {Morey, Leslie C. and Agresti, Alan},
  title   = {The measurement of classification agreement: An adjustment to the {Rand} statistic for chance agreement},
  journal = {Educational and Psychological Measurement},
  year    = {1984},
  volume  = {44},
  number  = {1},
  pages   = {33--37},
  doi     = {10.1177/0013164484441004}
}

@article{rand1971objective,
  author  = {Rand, William M.},
  title   = {Objective criteria for the evaluation of clustering methods},
  journal = {Journal of the American Statistical Association},
  year    = {1971},
  volume  = {66},
  number  = {336},
  pages   = {846--850},
  doi     = {10.1080/01621459.1971.10482356}
}

@article{held2006bayesian,
  author  = {Held, Leonhard and Holmes, Chris C.},
  title   = {Bayesian auxiliary variable models for binary and multinomial regression},
  journal = {Bayesian Analysis},
  year    = {2006},
  volume  = {1},
  number  = {1},
  pages   = {145--168},
  doi     = {10.1214/06-BA105}
}

@article{polson2013bayesian,
  author  = {Polson, Nicholas G. and Scott, James G. and Windle, Jesse},
  title   = {Bayesian inference for logistic models using {P{\'o}lya--Gamma} latent variables},
  journal = {Journal of the American Statistical Association},
  year    = {2013},
  volume  = {108},
  number  = {504},
  pages   = {1339--1349},
  doi     = {10.1080/01621459.2013.829001}
}

@article{chakraborty2024gibbs,
  title   = {A {Gibbs} posterior framework for fair clustering},
  author  = {Chakraborty, Abhisek and Bhattacharya, Anirban and Pati, Debdeep},
  journal = {Entropy},
  year    = {2024},
  volume  = {26},
  number  = {1},
  pages   = {63},
  doi     = {10.3390/e26010063}
}

@article{bissiri2016general,
  author  = {Bissiri, Pier Giovanni and Holmes, Chris C. and Walker, Stephen G.},
  title   = {A general framework for updating belief distributions},
  journal = {Journal of the Royal Statistical Society: Series B (Statistical Methodology)},
  year    = {2016},
  volume  = {78},
  number  = {5},
  pages   = {1103--1130},
  doi     = {10.1111/rssb.12158}
}

@article{rigon2023generalized,
  title={A generalized {Bayes} framework for probabilistic clustering},
  author={Rigon, Tommaso and Herring, Amy H and Dunson, David B},
  journal={Biometrika},
  volume={110},
  number={3},
  pages={559--578},
  year={2023},
  publisher={Oxford University Press}
}

@article{okabe2001replica,
  author  = {Okabe, Tsuneyasu and Kawata, Masaaki and Okamoto, Yuko and Mikami, Masuhiro},
  title   = {Replica-exchange {Monte Carlo} method for the isobaric--isothermal ensemble},
  journal = {Chemical Physics Letters},
  year    = {2001},
  volume  = {335},
  number  = {5--6},
  pages   = {435--439},
  doi     = {10.1016/S0009-2614(00)01309-3}
}

@inproceedings{geyer1991markov,
  title     = {Markov chain {Monte Carlo} maximum likelihood},
  author    = {Geyer, Charles J.},
  booktitle = {Computing Science and Statistics: Proceedings of the 23rd Symposium on the Interface},
  pages     = {156--163},
  year      = {1991},
  publisher = {Interface Foundation of North America},
  address   = {Fairfax Station, VA}
}

@article{syed2022non,
  author  = {Syed, Saifuddin and Bouchard-C{\^o}t{\'e}, Alexandre and Deligiannidis, George and Doucet, Arnaud},
  title   = {Non-reversible parallel tempering: a scalable highly parallel {MCMC} scheme},
  journal = {Journal of the Royal Statistical Society: Series B (Statistical Methodology)},
  year    = {2022},
  volume  = {84},
  number  = {2},
  pages   = {321--350},
  doi     = {10.1111/rssb.12477}
}

@article{fryzlewicz2014wild,
  title   = {Wild binary segmentation for multiple change-point detection},
  author  = {Fryzlewicz, Piotr},
  journal = {The Annals of Statistics},
  volume  = {42},
  number  = {6},
  pages   = {2243--2281},
  year    = {2014},
  doi     = {10.1214/14-AOS1245},
}

@article{kovacs2023seeded,
  title={Seeded binary segmentation: a general methodology for fast and optimal changepoint detection},
  author={Kov{\'a}cs, Solt and B{\"u}hlmann, Peter and Li, Housen and Munk, Axel},
  journal={Biometrika},
  volume={110},
  number={1},
  pages={249--256},
  year={2023},
  publisher={Oxford University Press}
}

@article{korkas2022ensemble,
  title={Ensemble binary segmentation for irregularly spaced data with change-points},
  author={Korkas, Karolos K},
  journal={Journal of the Korean Statistical Society},
  volume={51},
  number={1},
  pages={65--86},
  year={2022},
  publisher={Springer}
}

@article{zhang2021graph,
  author  = {Zhang, Yuxuan and Chen, Hao},
  title   = {Graph-based multiple change-point detection},
  journal = {arXiv preprint arXiv:2110.01170},
  year    = {2021}
}

@article{ross2021nonparametric,
  author  = {Ross, Gordon J.},
  title   = {Nonparametric detection of multiple location-scale change points via wild binary segmentation},
  journal = {arXiv preprint arXiv:2107.01742},
  year    = {2021}
}

@article{truong2020selective,
  title={Selective review of offline change point detection methods},
  author={Truong, Charles and Oudre, Laurent and Vayatis, Nicolas},
  journal={Signal Processing},
  volume={167},
  pages={107299},
  year={2020},
  publisher={Elsevier}
}

@article{carvalho2010horseshoe,
  title={The horseshoe estimator for sparse signals},
  author={Carvalho, Carlos M and Polson, Nicholas G and Scott, James G},
  journal={Biometrika},
  volume={97},
  number={2},
  pages={465--480},
  year={2010}
}

@article{makalic2015simple,
  title={A simple sampler for the horseshoe estimator},
  author={Makalic, Enes and Schmidt, Daniel F},
  journal={IEEE Signal Processing Letters},
  volume={23},
  number={1},
  pages={179--182},
  year={2015},
  publisher={IEEE}
}

@article{bhattacharyya2022applications,
  author  = {Bhattacharyya, Arinjita and Pal, Subhadip and Mitra, Riten and Rai, Shesh},
  title   = {Applications of {Bayesian} shrinkage prior models in clinical research with categorical responses},
  journal = {BMC Medical Research Methodology},
  year    = {2022},
  volume  = {22},
  pages   = {126},
  doi     = {10.1186/s12874-022-01605-0}
}

@article{peterson1954theory,
  author  = {Peterson, W. W. and Birdsall, T. G. and Fox, W. C.},
  title   = {The theory of signal detectability},
  journal = {Transactions of the IRE Professional Group on Information Theory},
  year    = {1954},
  volume  = {4},
  number  = {4},
  pages   = {171--212},
  doi     = {10.1109/TIT.1954.1057460}
}

@article{robin2011proc,
  author  = {Robin, Xavier and Turck, Natacha and Hainard, Alexandre and Tiberti, Natalia and Lisacek, Fr{\'e}d{\'e}rique and Sanchez, Jean-Charles and M{\"u}ller, Markus},
  title   = {{pROC}: an open-source package for {R} and {S+} to analyze and compare {ROC} curves},
  journal = {BMC Bioinformatics},
  year    = {2011},
  volume  = {12},
  pages   = {77},
  doi     = {10.1186/1471-2105-12-77}
}

@article{oh2008learning,
  title={Learning and inferring motion patterns using parametric segmental switching linear dynamic systems},
  author={Oh, Sang Min and Rehg, James M and Balch, Tucker and Dellaert, Frank},
  journal={International Journal of Computer Vision},
  volume={77},
  number={1},
  pages={103--124},
  year={2008},
  publisher={Springer}
}

@article{candanedo2016accurate,
  author  = {Candanedo, Luis M. and Feldheim, V{\'e}ronique},
  title   = {Accurate occupancy detection of an office room from light, temperature, humidity and {CO$_2$} measurements using statistical learning models},
  journal = {Energy and Buildings},
  year    = {2016},
  volume  = {112},
  pages   = {28--39},
  doi     = {10.1016/j.enbuild.2015.11.071}
}

@article{kei2023change,
  author  = {Kei, Yik Lun and Li, Hangjian and Chen, Yanzhen and Madrid Padilla, Oscar Hernan},
  title   = {Change Point Detection on A Separable Model for Dynamic Networks},
  journal = {Transactions on Machine Learning Research},
  year    = {2025},
  url     = {https://openreview.net/forum?id=DSNJykzHF3}
}

@article{
crozier2025,
author = {Peter A. Crozier  and Matan Leibovich  and Piyush Haluai  and Mai Tan  and Andrew M. Thomas  and Joshua Vincent  and Sreyas Mohan  and Adria Marcos Morales  and Shreyas A. Kulkarni  and David S. Matteson  and Yifan Wang  and Carlos Fernandez-Granda },
title = {Visualizing nanoparticle surface dynamics and instabilities enabled by deep denoising},
journal = {Science},
volume = {387},
number = {6737},
pages = {949-954},
year = {2025},
doi = {10.1126/science.ads2688},
URL = {https://www.science.org/doi/abs/10.1126/science.ads2688},
eprint = {https://www.science.org/doi/pdf/10.1126/science.ads2688}
}

@article{moon2023,
author = {Chul Moon and Qiwei Li and Guanghua Xiao},
title = {{Using persistent homology topological features to characterize medical images: Case studies on lung and brain cancers}},
volume = {17},
journal = {The Annals of Applied Statistics},
number = {3},
publisher = {Institute of Mathematical Statistics},
pages = {2192 -- 2211},
year = {2023},
doi = {10.1214/22-AOAS1714},
}

@article{vincent2021,
  author  = {Vincent, Joshua L. and Crozier, Peter A.},
  title   = {Atomic level fluxional behavior and activity of {CeO$_2$}-supported {Pt} catalysts for {CO} oxidation},
  journal = {Nature Communications},
  year    = {2021},
  volume  = {12},
  number  = {1},
  pages   = {5789},
  doi     = {10.1038/s41467-021-26047-8}
}

@article{hough_transform,
  title={Use of the {Hough} transformation to detect lines and curves in pictures},
  author={Duda, Richard O and Hart, Peter E},
  journal={Communications of the ACM},
  volume={15},
  number={1},
  pages={11--15},
  year={1972},
  publisher={ACM New York, NY, USA}
}

@article{li1993,
  author  = {Li, Chun Hung and Lee, C. K.},
  title   = {Minimum cross entropy thresholding},
  journal = {Pattern Recognition},
  year    = {1993},
  volume  = {26},
  number  = {4},
  pages   = {617--625}
}

@inproceedings{garin2019,
  author    = {Garin, Ad{\'e}lie and Tauzin, Guillaume},
  title     = {A topological ``reading'' lesson: Classification of {MNIST} using {TDA}},
  booktitle = {2019 18th IEEE International Conference on Machine Learning and Applications (ICMLA)},
  year      = {2019},
  pages     = {1551--1556},
  publisher = {IEEE},
  doi       = {10.1109/ICMLA.2019.00253}
}

@article{thomas2023,
  title={Feature detection and hypothesis testing for extremely noisy nanoparticle images using topological data analysis},
  author={Thomas, Andrew M and Crozier, Peter A and Xu, Yuchen and Matteson, David S},
  journal={Technometrics},
  volume={65},
  number={4},
  pages={590--603},
  year={2023},
  publisher={Taylor \& Francis}
}

@article{green1995reversible,
  title={Reversible jump {Markov} chain {Monte Carlo} computation and {Bayesian} model determination},
  author={Green, Peter J},
  journal={Biometrika},
  volume={82},
  number={4},
  pages={711--732},
  year={1995},
  publisher={Oxford University Press}
}

@book{Chen2012,
  author    = {Chen, Jie and Gupta, Arjun K.},
  title     = {Parametric Statistical Change Point Analysis},
  publisher = {Birkh{\"a}user},
  address   = {Boston, MA},
  year      = {2012},
  edition   = {2nd}
}

@article{longin2001extreme,
  author  = {Longin, Fran{\c{c}}ois and Solnik, Bruno},
  title   = {Extreme correlation of international equity markets},
  journal = {The Journal of Finance},
  year    = {2001},
  volume  = {56},
  number  = {2},
  pages   = {649--676},
  doi     = {10.1111/0022-1082.00340}
}
\newpage
    \begin{center}
    \Large \textbf{Supplementary Material for \\
    ``A generalized Bayesian approach to multiple changepoint analysis"}
\end{center}

\section{Implementation details in simulation studies} \label{Supp. implementation details}
This section explains how we obtained the simulation results in Section~\ref{Sec. simualtion results} of the main article, including how we implemented the competing methods and the proposed \texttt{bcmlr} method in Section~\ref{Sec. bcmlr model} and computed the adjusted Rand index. 

\paragraph{Implementation of \texttt{bcmlr}. } When the number of changepoints is known, we estimated the changepoint locations by fitting two changepoints using the \texttt{bcmlr} Gibbs sampler in Section~\ref{Sec. bcmlr Gibbs sampler} with $m = 30$. When the number of changepoints is unknown, we estimated the number and locations of changepoints using the approach in Section~\ref{Sec. select num of CPs} with these tuning parameters: $\tau = 0.5$, $\alpha = 0.1$, $\zeta = 5$, $m = 30$, and $L_{\text{fitted}} = 5$. The Gibbs samplers were initialized at positions that evenly split the series and run $5{,}000$ iterations, with the first $2{,}500$ discarded as burn-in. Since changes occur in relatively few dimensions compared with the total number of dimensions based on the data generating processes, we assigned horseshoe priors for the regression coefficients under all scenarios. 

\paragraph{Implementation of \texttt{kcp} and \texttt{MultiRank}. } When the number of changepoints is known, the \texttt{ruptures} package \citep{truong2020selective} provides functions that implement dynamic programming \citep{bellman1954theory} to get changepoint estimates from \texttt{MultiRank} and \texttt{kcp}. A radial basis function \texttt{CostRbf} is used for \texttt{kcp}, while a rank-based cost function \texttt{CostRank} is used for \texttt{MultiRank}. For both methods, the minimum segment length is set to 30 (same as for \texttt{bcmlr} and \texttt{e.divisive}). When the number of changepoints is unknown, we use the functions implementing the ``slope heuristics" method \citep{arlot2019kernel} available in \citet{londschien2023random}'s GitHub simulation repository. The implementation of \texttt{kcp} and \texttt{MultiRank} requires specifying a maximum number of changepoints. We set the maximum number of changepoints equal to 5. 

\paragraph{Implementation of \texttt{e.divisive}. } The \texttt{e.divisive} function is implemented in the \texttt{ecp} package \citep{matteson2014nonparametric} and includes an input argument $k$ that specifies the number of changepoints. When the number of changepoints is known, we set $k = 2$, in which case \texttt{e.divisive} performs exactly $k$ bisections to identify $k$ changepoint locations. When the number of changepoints is unknown, we do not specify a value for $k$. In this setting, \texttt{e.divisive} recursively bisects the series and, after each bisection, applies a statistical test based on a multivariate divergence measure to assess the significance of a changepoint. The implementation of \texttt{e.divisive} also requires specifying a minimum segment length. Regardless of whether the number of changepoints is known or unknown, we set the minimum segment length to 30, which is the default value recommended by \citet{matteson2014nonparametric}.

\paragraph{Implementation of \texttt{ClaSP}. } The \texttt{ClaSPy} package \citep{ermshaus2023clasp} provides the \texttt{BinaryClaSPSegmentation} function for implementing the \texttt{ClaSP} method. Depending on whether the number of changepoints is known, two parameters—the number of segments and the use of validation—are specified differently. When the number of changepoints is known, we set the number of segments equal to 3 and disable validation. When the number of changepoints is unknown, we do not specify the number of segments and instead enable validation with a classification score threshold of 0.75. These hyperparameter settings are recommended by the authors of \texttt{ClaSP} in their GitHub documentation.

\paragraph{Implementation of \texttt{changeforest}. } The \texttt{changeforest} package \citep{londschien2023random}, available in both R and Python, provides the \texttt{changeforest} function. This function does not include a parameter for specifying the number of changepoints. In our implementation, we set the classifier to a random forest and the search method to binary segmentation, which are the default settings of the function.

\paragraph{Computation of the adjusted Rand index.} To compute the ARI in R, we used the \texttt{adj.rand.index} function from the \texttt{pdfCluster} package \citep{azzalini2014clustering}. This function takes as input two vectors of equal length, each containing class labels corresponding to a particular partition of the data. In the context of changepoint detection, these vectors are constructed from the class labels induced by the true and estimated changepoint locations. For evaluating changepoint detection methods implemented in Python, the \texttt{sklearn.metrics} package \citep{pedregosa2011scikit} provides a function to compute the ARI, which similarly takes the true and predicted labels as inputs.

\section{Additional data applications}\label{Supp. Additional real data applications}

As mentioned in Section~\ref{Sec. real data app} of the main article, we apply \texttt{bcmlr} to multivariate series with annotations provided by \citet{van2020evaluation}. The analysis on the annotated series serves as a middle ground between simulations studies (in Section~\ref{Sec. Simulation} of the main article) with known ground truth and real data applications without knowledge of the true changepoints in Section~\ref{Sec. real data app}. 

\subsection{Honey bee dance}\label{Sec. honey bee app}



\begin{figure}[h]
    \centering
     \includegraphics[width=\textwidth]{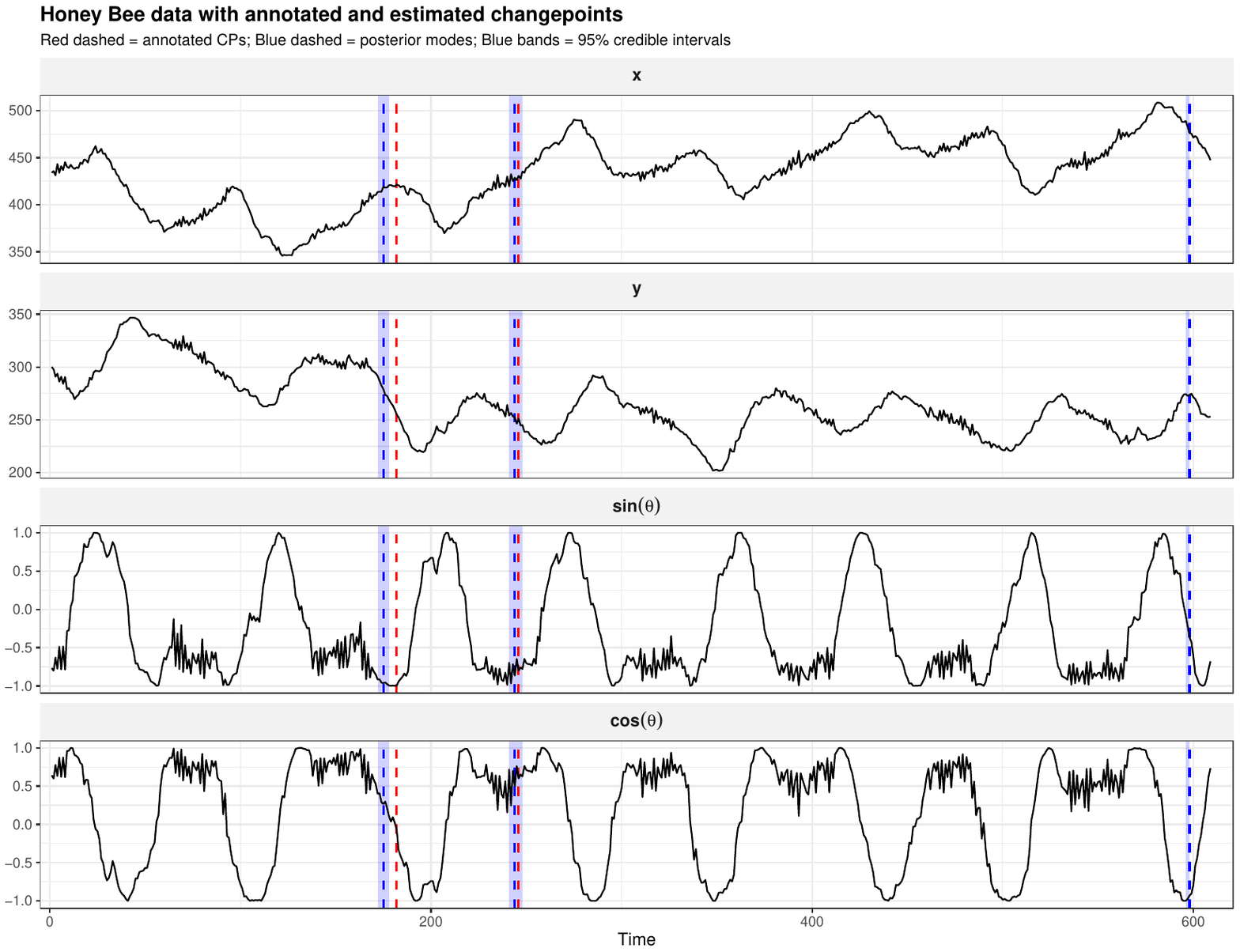}
    \caption{A time series representation of a honey bee dance trajectory with annotated changepoints (red dashed lines) and posterior mode estimates of changepoints from $\texttt{bcmlr}$ (blue dashed lines). The blue bands are 95\% credible intervals. }
    \label{fig:bee-time-series}
\end{figure}

\citet{van2020evaluation} provided annotations for one of the dance trajectories of length 609 originally provided by \citet{oh2008learning}. The honey been dance trajectory is represented by 4-dimensional vectors: $z_t = \left[x_t, y_t, \operatorname{sin}(\theta_t), \operatorname{cos}(\theta_t) \right]$, where $x_t, y_t$ are the 2D coordinates, and $\theta_t$ is the dancer bee's heading angle at time $t$ (Figure~\ref{fig:bee-time-series}). On average, the number of changepoints annotated was 0.4: Four out of five human annotators believed there were no changepoints, while one annotator identified two changepoints: 182 and 247. 

Since the true number of changepoints $L_{\text{true}}$ is unknown in the honey bee dance data, we apply the approach to selecting the number of changepoints described in Section \ref{Sec. select num of CPs}. Based on the annotator’s low confidence in the presence of changepoints in this series, we set the number of fitted changepoints to $L_{\text{fitted}} = 4$ and choose a confidence level of $1-\alpha = 0.95$ for the confidence interval for the AUC. In addition, we set the AUC threshold $\tau = 0.5$, minimum segment length $m = 10$, and hold out every 5th ($\zeta = 5$) observation for AUC computation. Since the data is relatively low-dimensional, we use a multivariate Gaussian distribution $\mathcal{N}(\bm{0}, 3\bm{I})$ as the prior distribution for the regression coefficients. We center and standardize the series before applying our method. We initialize the changepoints at evenly spaced positions along the series and run the Gibbs sampler for $T = 15{,}000$ iterations, with a burn-in period of $T_0 = 10{,}000$ iterations. We obtain posterior samples of the number of fitted changepoints that correspond to true changepoints, $L_{\text{true}}^{(1)}, \dots, L_{\text{true}}^{(5000)}$, and the posterior mode is $\hat{L}_{\text{true}} = 3$. 

We reapply the $\texttt{bcmlr}$ method on all the data to fit 3 changepoints and find that the first two of the estimated changepoints, $\kappa_1 = 175$, $\kappa_2 = 244$, and $\kappa_3 = 598$, closely align with the annotated ones, 182 and 247. Figure~\ref{fig:bee-time-series} presents the annotated changepoints (red dashed lines) alongside the estimated changepoints (blue dashed lines) and 95\% credible intervals (blue bands).

\begin{figure}[h]
    \centering
    \includegraphics[width= \textwidth]{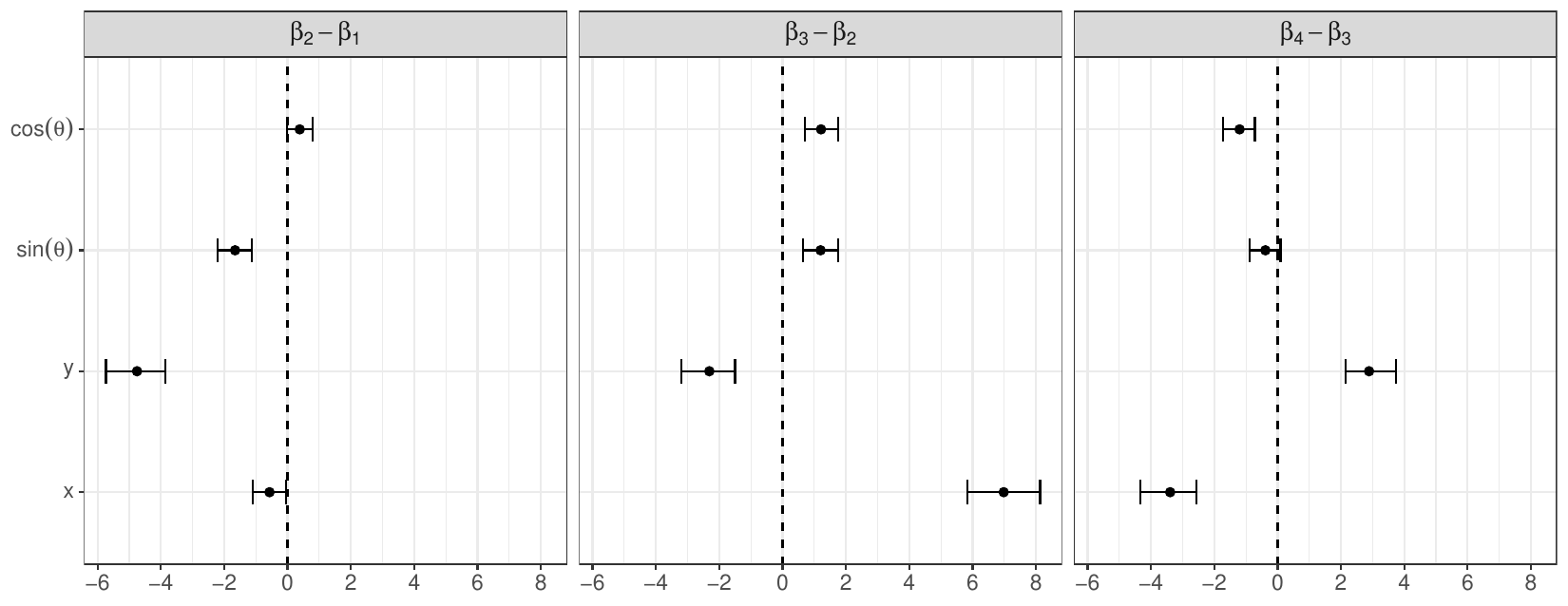}
    \caption{Each plot shows credible intervals of the differences between the coefficient vectors
  $\bm{\beta}_{l+1}-\bm{\beta}_{l}$, for $ l \in \{1, \dots, 3\}$,  with 4 dimensions
  corresponding to $x, y, \operatorname{sin}(\theta)$ and $\operatorname{cos}(\theta)$. The black dots represent posterior mean coefficient values.}
    \label{fig:bee-CIs}
\end{figure}

Using posterior summaries of the coefficients in the multinomial logistic regression implied by our loss function, we can interpret which covariates contributed to the changes. In particular, we can look at 95\% credible intervals of $\bm{\beta}_{l+1} -\bm{\beta}_{l}$, the differences between posterior coefficients that distinguish data in neighboring segments on either side of the estimated changepoint $\kappa_l$ (Figure~\ref{fig:bee-CIs}). We can interpret the estimated changepoints in the same manner as in Section~\ref{Sec. real data app} of the main article. For example, in the first subplot, the credible intervals corresponding to sin$(\theta)$, $x$, and $y$ are in the negative ranges, while the credible interval corresponding to cos$(\theta)$ is in the positive range. Thus, $\kappa_1 = 182$ corresponds to the increase of $\text{cos}(\theta)$ and the decrease of $\text{sin}(\theta), x$ and $y$ in the honey bee dance trajectory. Similarly, the remaining subplots provide insight into what changes occurred around the other detected changepoints. By looking at the differences between posterior coefficients, we are able to provide intuitive interpretation about subtle changepoints in a honey bee dance trajectory (Figure~\ref{fig:bee-time-series}). 

The competing methods identify different changepoints in the honey bee dance series, while the changepoints detected by our method are closer to the annotated changepoints. The \texttt{e.divisive} method seems to flag noise and overestimate the number of changepoints in the honey bee dance series when the minimum segment length is not chosen carefully. For example, \texttt{e.divisive} (with a minimum segment length equal to 10) detects 48 changepoints. However, with a minimum segment length to 30, \texttt{e.divisive} detects 15 changepoints, located at 42, 87, 119, 174, 204, 235, 269, 299, 333, 363, 402, 440, 472, 503, and 558. In contrast, \texttt{kcp}, \texttt{MultiRank}, \texttt{ClaSP} and \texttt{changeforest} are conservative regarding the existence of changepoints in this series, which aligns with the opinions of most human annotators. \texttt{kcp} (with a maximum number of changepoints equal to 4) does not identify a changepoint whereas \texttt{MultiRank} detects one changepoint at 176. Neither \texttt{ClaSP} nor \texttt{changeforest} detects any changepoints in this series. 


\subsection{Room Occupancy Detection}

\begin{figure}[h]
    \centering
    \includegraphics[width=\textwidth]{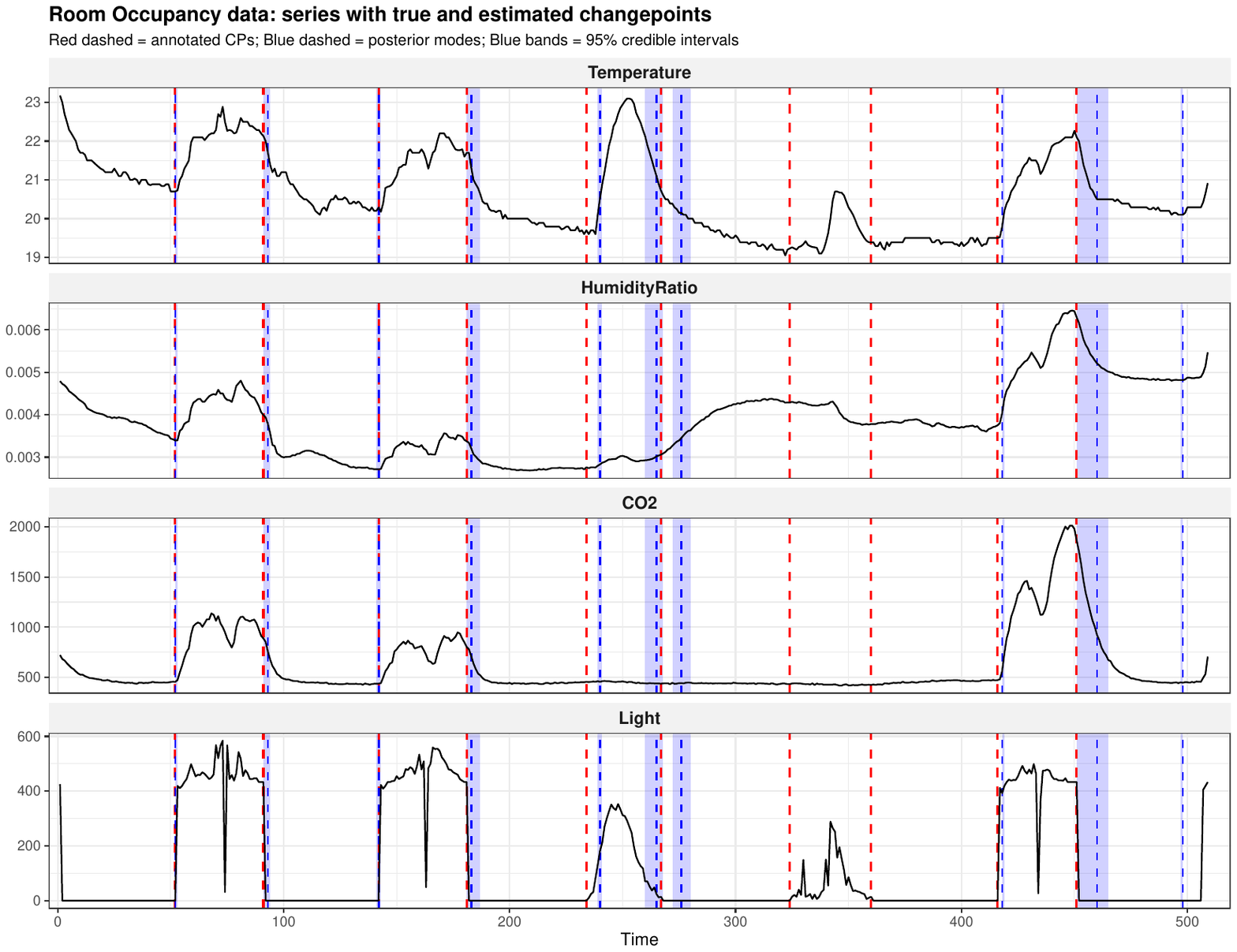}
    \caption{The room occupancy time series with annotated changepoints (red dashed lines) and estimated changepoints from $\texttt{bcmlr}$ (blue dashed lines). The blue bands are 95\% credible intervals. }
    \label{fig:room-occupancy-time-series}
\end{figure}

This room occupancy time series (Figure \ref{fig:room-occupancy-time-series}) is extracted from \citet{candanedo2016accurate}'s training series at every 16 minutes and has 509 observations of light, temperature, humidity, and $\text{CO}_2$. According to \citet{van2020evaluation}, all five human annotators identified different numbers of changepoints ranging from 2 to 10, yielding an average of 6.2 changepoints. The most inclusive annotation includes 10 changepoints: 52, 91, 142, 181, 234, 267, 324, 360, 416, 451. 

Since the true number of changepoints $L_{\text{true}}$ is unknown for the room occupancy data, we apply the approach to selecting the number of changepoints described in Section \ref{Sec. select num of CPs}. Based on the annotations, we set the number of fitted changepoints to $L_{\text{fitted}} = 10$ and choose a confidence level of $1-\alpha = 0.95$ for the confidence interval for the AUC. In addition, we set the AUC threshold $\tau = 0.5$, minimum segment length $m = 10$, and hold out every 5th ($\zeta = 5$) for AUC computation. Since the series is relatively low-dimensional, we use a multivariate Gaussian distribution $\mathcal{N}(\bm{0}, 3\bm{I})$ for the regression coefficients. We center and standardize the series before applying our method. To initialize the Gibbs sampler, we position the changepoints at evenly spaced locations along the series. We run the Gibbs sampler for $T = 15{,}000$ iterations, with a burn-in of $T_0 = 10{,}000$ iterations. We obtain posterior samples of the number of fitted changepoints that correspond to true changepoints, $L_{\text{true}}^{(1)}, \dots, L_{\text{true}}^{(5000)}$, and the posterior mode is $\hat{L}_{\text{true}} = 10$. 

We reapply the $\texttt{bcmlr}$ method to all the data to fit 10 changepoints and find 8 out of 10 of the estimated changepoints closely align with the annotated changepoints. The estimated changepoints are $\kappa_1 = 52$, $\kappa_2 = 93$, $\kappa_3 = 142$, $\kappa_4 = 183$, $\kappa_5 = 240$, $\kappa_6 = 265$, $\kappa_7 = 276$, $\kappa_8 = 418$, $\kappa_9 = 460$, and $\kappa_{10} = 498$. Figure \ref{fig:room-occupancy-time-series} displays the annotated changepoints (red dashed lines) alongside the estimated changepoints (blue dashed lines) and 95\% credible intervals (blue bands).

\begin{figure}[h]
    \centering
    \includegraphics[width= \textwidth]{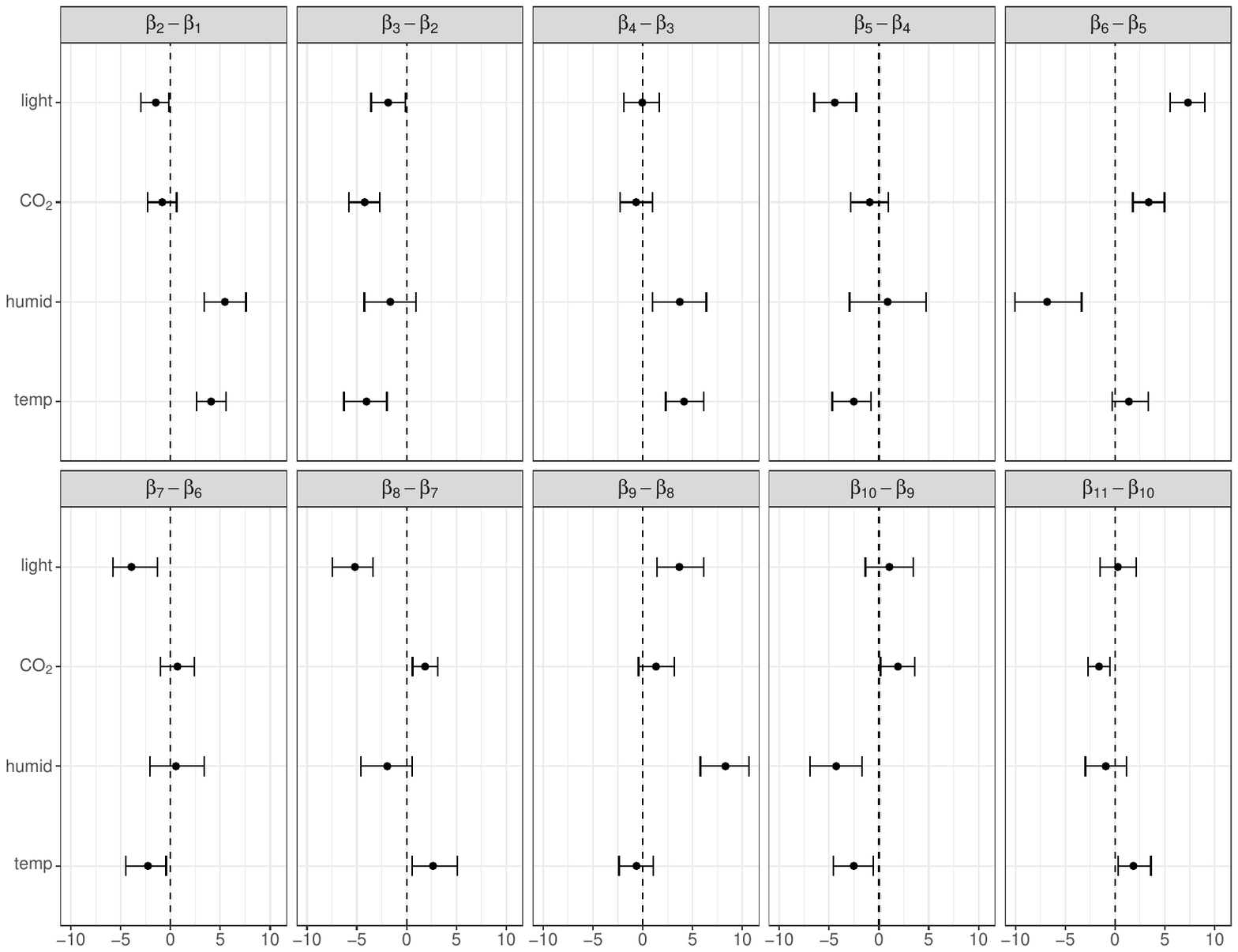}
    \caption{Each plot shows credible intervals of the differences between the coefficient vectors
  $\bm{\beta}_{l+1}-\bm{\beta}_{l}$, for $ l \in \{1, \dots, 10\}$, with 4 dimensions
  corresponding to temperature, humidity ratio, $\text{CO}2$, and light. The black dots represent posterior mean coefficient values.}
    \label{fig:room-occupancy-CIs}
\end{figure}

Using posterior summaries of the coefficients in the multinomial logistic regression implied by our loss function, we can interpret which covariates contributed to the changes. In particular, we look at 95\% credible intervals of the differences between $\bm{\beta}_{l+1} -\bm{\beta}_{l}$, as shown in Figure~\ref{fig:room-occupancy-CIs}. We can interpret the estimated changepoints in the same manner as in Section~\ref{Sec. real data app} of the main article. For example, in the first subplot, the credible intervals corresponding to the humidity ratio and temperature are in the positive ranges, the credible interval corresponding to light lies below zero, while the credible interval corresponding to CO$_2$ includes zero. Thus, $\kappa_1 = 52$ corresponds to the increase of the humidity ratio and temperature and the decrease of CO$_2$ in the room. A practical scenario with these changes is when people leave the room and turn off the air conditioning, thereby heating and humidifying the room while reducing CO$_2$ levels. Similarly, one can interpret the other detected changepoints using the rest of the subplots. 

Compared with the competing methods, our method produces estimated changepoints that are closer to the annotations. The \texttt{e.divisive} and \texttt{changeforest} methods seem to overestimate the number of changepoints, while the \texttt{kcp} and \texttt{MultiRank} methods are conservative about the existence of changepoints. \texttt{e.divisive} (with a minimum segment length equal to 10) detects 22 changepoints: 12, 54, 64, 92, 102, 143, 153, 170, 182, 192, 240, 258, 268, 340, 351, 369, 411, 421, 441, 460, 470, 480, where half of the estimates are close to the annotations. \texttt{changeforest} detects 59 changepoints. \texttt{kcp} (with a maximum number of changepoints equal to 10) does not detect any changepoints, while \texttt{MultiRank} detects 2 changepoints: 280, 370. \texttt{ClaSP} detects no changepoints in this series.


\section{Parallel tempering for changepoint detection} \label{Supp. Parallel tempering}

As mentioned in Section~\ref{Sec. Accelerating mixing}, we can use parallel tempering to improve the mixing of the Gibbs sampler for some very challenging configurations of the true changepoints. Parallel tempering was originally proposed by \citet{geyer1991markov}. Since then, the method has been extended and applied in many fields to facilitate sampling from multimodal target distributions.

To implement parallel tempering for our generalized Bayesian method in Section~\ref{Sec. bcmlr model}, we need the tempered versions of the posterior density in Section~\ref{Sec. bcmlr model}, Gibbs samplers to sample from the tempered versions of the augmented posterior density, and the acceptance ratio for Metropolis-Hastings checks. 

First, we raise the exponential term of the negative loss to a tempering power $t \in (0, 1]$ to get a tempered version of the posterior density: 
\begin{align} \label{Eq. tempered bcmlr posterior}
     \pi_t \left(\bm{\kappa}, \bm{\beta}_1, \dots, \bm{\beta}_J | \bm{X}\right) \propto \operatorname{exp}\left\{ - t\cdot l\left(\bm{\kappa}, \bm{\beta}_1, \dots, \bm{\beta}_J, \bm{X} \right)\right\} \pi\left(\bm{\kappa}, \bm{\beta}_1, \dots, \bm{\beta}_J\right),
\end{align}
where 
\begin{align}
    t \cdot l\left(\boldsymbol{\kappa}, \bm{\beta}_1, \dots, \bm{\beta}_J, \boldsymbol{X}\right) 
    = - \log \left\{\left(\prod_{i=1}^N \prod_{j=1}^J \frac{e^{\eta_{ij}y_{ij}}}{1 + e^{\eta_{ij}}}\right)^t \right\}. \nonumber
\end{align}

To simulate from the tempered target posterior density \eqref{Eq. tempered bcmlr posterior}, we apply P\'olya-Gamma data augmentation and obtain an augmented posterior density (denoted as $\pi_t $) that yields tractable full conditional distributions \eqref{Eq. tempered GS - omega}, \eqref{Eq. tempered GS - beta}, and \eqref{Eq. tempered GS - kappa} as below. 
\begin{align}\label{Eq. tempered GS - omega}
    (\omega_{ij} \mid \bm{\beta}_1, \dots, \bm{\beta}_J, \boldsymbol{x}_i) \sim \operatorname{PG}\left(t, \eta_{ij} \right), \text{ for } i = 1, \dots, N;
\end{align}
\begin{align} \label{Eq. tempered GS - kappa}
 \pi_t\left(\kappa_l \mid \kappa_{l-1}, \kappa_{l+1}, \boldsymbol{B}, \boldsymbol{X}\right) \propto \prod_{j=l}^{l+1} \alpha_j^{\kappa_j -\kappa_{j-1}} \left(\prod_{i=\kappa_{l-1}+1}^{\kappa_l} \frac{e^{\boldsymbol{x}_i^{\top}\boldsymbol{\beta}_l}}{\sum_{k=1}^J e^{\boldsymbol{x}_i^{\top}\boldsymbol{\beta}_k}} \prod_{i = \kappa_l+1}^{\kappa_{l+1}} \frac{e^{\boldsymbol{x}_i^{\top}\boldsymbol{\beta}_{l+1}}}{\sum_{k=1}^J e^{\boldsymbol{x}_i^{\top}\boldsymbol{\beta}_k}}\right)^t; 
\end{align}
\begin{align}\label{Eq. tempered GS - beta}
    (\boldsymbol{\beta}_j \mid \boldsymbol{\beta}_{-j}, \boldsymbol{\omega}_j, \boldsymbol{y}_j, \boldsymbol{X}) \sim \boldsymbol{N}(\boldsymbol{m}_j, \boldsymbol{V}_j),
\end{align}
where 
$$
\begin{aligned}
    \boldsymbol{V}_j & = (\boldsymbol{X}^{\top}\boldsymbol{\Omega}_j \boldsymbol{X} + \boldsymbol{\Sigma}_{0j}^{-1})^{-1},\\
    \boldsymbol{m}_j & = \boldsymbol{V}_j\left(\boldsymbol{X}^{\top}(\boldsymbol{\Omega}_j \boldsymbol{c}_j + \boldsymbol{\delta}_j) + \boldsymbol{\Sigma}_{0j}^{-1} \boldsymbol{\mu}_{0j}\right), \\
    \boldsymbol{\Omega}_j & = \operatorname{diag}(\boldsymbol{\omega}_j), \quad 
    \boldsymbol{\delta}_j = t\left(\boldsymbol{y}_j - \mathbbm{1}_N \cdot \frac{1}{2}\right). 
\end{aligned}
$$
These full conditionals are derived under Gaussian priors for regression coefficients $\bm{\beta}_j \sim \mathcal{N}\left(\bm{\mu}_{0j}, \bm{\Sigma}_{0j}\right)$ and the segment-length-based prior for changepoints \eqref{Eq. bcmlr kappa prior}.

The communication among the chains is determined by Metropolis-Hastings checks. We denote the samples from the augmented posterior density as $\bm{\theta}^t  = \left(\boldsymbol{\beta}^{t}, \boldsymbol{\omega}^{t}, \boldsymbol{\kappa}^{t}\right)$. For two tempering powers $t_1, t_2 \in (0,1]$ and $t_1 \neq t_2$, the samples $\bm{\theta}^{t_1}$ and $\bm{\theta}^{t_2}$ will be swapped if a random uniform variable $u\sim \operatorname{Uniform}(0,1)$ is less than  $\min\{1, A^{(t_1, t_2)}\}$, where 
\begin{align}\label{Eq. MH acceptance probability}
    A^{(t_1, t_2)} = \quad \frac{\left(\pi_{t_1} \text{ evaluated at } \boldsymbol{\theta}^{t_2} \right) \cdot \left(\pi_{t_2} \text{ evaluated at } \boldsymbol{\theta}^{t_1} \right)}{\left(\pi_{t_1} \text{ evaluated at } \boldsymbol{\theta}^{t_1} \right) \cdot \left(\pi_{t_2} \text{ evaluated at } \boldsymbol{\theta}^{t_2} \right) }. 
\end{align}

Two common ways to select a pair of samples are the deterministic even-odd (DEO) scheme \citep{okabe2001replica}, where samples with even and odd indices are chosen in an alternating, non-reversible fashion, and the stochastic even-odd (SEO) scheme, where samples with even and odd indices are chosen randomly. \citet{syed2022non} showed that DEO outperforms SEO. 

To achieve thorough mixing across modes using a parallel tempering algorithm and obtain samples representative of the target posterior distribution, it is useful to find the optimal tempering schedule, i.e., the optimal set of tempering powers. To find the optimal schedule, \citet{syed2022non} proposed  estimating a communication barrier function $\Lambda$ and solving a minimization problem based on the estimated $\Lambda$. A communication barrier at a tempering power $t$ approximates the sum of Metropolis-Hastings rejection probabilities from the first tempering power up to $t$. A low value of $\Lambda$ indicates low rejection rates and more active swaps between the chains at different tempering powers. 

\citet{syed2022non}'s non-reversible parallel tempering method, presented in Algorithm 4 of their article, starts with finding the optimal tempering schedule. First, they initialize a tempering schedule. For a selected number of iterations, they perform steps i) to iii) to estimate the length of the optimal tempering schedule: i) Obtain a list of Metropolis-Hastings rejection probabilities under the DEO scheme (Algorithm 1). ii) Compute a communication barrier $\hat{\Lambda}(t)$ under every tempering power $t$. Estimate a tempering schedule by fitting a monotone increasing interpolation $\hat{\Lambda}(\cdot)$ based on all tempering powers and the corresponding communication barriers (Algorithm 3). iii) Solve optimization problems based on $\hat{\Lambda}$ (Algorithm 2) to obtain the estimated tempering powers. After iterating these steps to tune the communication barrier and tempering schedules, the optimal number of tempering powers equals twice the value of the latest communication barrier function evaluated at tempering power 1. Then, they apply Algorithm 2 using the latest communication barrier function and the optimal number of tempering powers to find the optimal tempering schedule. Finally, using Algorithm 1, they obtain posterior samples from all tempered versions of the target function under the optimal tempering schedule. 

To simulate from our posterior distribution in Section \ref{Sec. bcmlr model} using non-reversible parallel tempering, we set the tempered target functions, which are called ``local exploration kernels" by \citet{syed2022non}, to be the tempered posterior density functions $\pi_t$ \eqref{Eq. tempered bcmlr posterior} and use the Gibbs sampler consisting of steps \eqref{Eq. tempered GS - omega} to \eqref{Eq. tempered GS - beta} to update posterior samples. A complete function implementing our method with non-reversible parallel tempering is available in our GitHub repository.

\section{Alternative priors for \texttt{BCLR}}\label{Supp. BCLR Gibbs sampler under horseshoe}

In the main article, we reviewed \citet{thomas2024bayesian}'s generalized Bayesian method, \texttt{BCLR}, for single changepoint detection, and a Gibbs sampler they constructed under a multivariate Gaussian prior for the regression coefficients and a uniform prior for the changepoint. We consider different priors for the changepoint and the regression coefficients under the \texttt{BCLR} method and present another Gibbs sampler for posterior simulation.

We choose a prior for $\kappa$ based on the lengths of the segments such that 
\begin{align}\label{Eq. Supp. bclr kappa prior}
    \pi(\kappa) \propto 
    \left(\frac{1}{\kappa} \right)^{\kappa} \left(\frac{1}{N-\kappa} \right)^{N-\kappa}. 
\end{align} This choice of prior for $\kappa$ penalizes small segments and assigns the greatest mass to the case when the two homogeneous segments are of equal length. 

When the number of coefficients $p$ is small, a multivariate Gaussian prior is a standard choice. When $p$ is large and we expect the changes to occur in only a few dimensions, we recommend using the horseshoe prior as in \eqref{Eq. HS prior}. Given the horseshoe prior, the full conditional distribution of $\bm{\beta}$ is 
$$
\begin{aligned}
    \bm{\beta} \mid \kappa, \bm{\omega}, \bm{\lambda}, \tau \sim \mathcal{N}\left(\bm{m}, \bm{V}\right), \quad \bm{\lambda} &= \left(\lambda_1, \dots, \lambda_p\right)\\
    \bm{V} = \left(\bm{X}^{\top}\bm{\Omega}\bm{X} + \bm{\Sigma}_0\right)^{-1}, \quad \bm{m} = \bm{V}\bm{X}^{\top}\bm{\delta}, \quad
    \bm{\Sigma}_0 &= \operatorname{diag}\left(\lambda_1^2\tau^2, \dots, \lambda_p^2\tau^2\right), \\
    \bm{\Omega} = \operatorname{diag}\left(\bm{\omega}\right), \quad \bm{\omega} = \left(\omega_1, \dots, \omega_N\right), \quad \bm{\delta} &= \bm{y} -  \mathbbm{1}_N \cdot \frac{1}{2}.
\end{aligned}
$$
The full conditional distributions of the hyperparameters $\lambda_d, \nu_d, \tau$ and $\xi$ in the horseshoe prior under logistic regression come from \citet{bhattacharyya2022applications}. Based on the discrete prior distribution \eqref{Eq. Supp. bclr kappa prior} for the changepoint and the horseshoe prior for the regression coefficients, the Gibbs sampler that samples from the augmented posterior distribution under the \texttt{BCLR} method iterates through the following steps:
$$
\begin{aligned}
    \kappa \mid \bm{\beta}, \bm{X} &\sim \pi\left(\kappa \mid \bm{\beta}, \bm{X} \right), 
    \\
    \omega_i \mid \bm{\beta}, \bm{x}_i 
    & \sim \operatorname{PG}\left(1, \bm{x}_i^{\top}\bm{\beta}\right), \\
    \bm{\beta} \mid \kappa, \bm{\omega}, \bm{\lambda}, \tau &\sim \mathcal{N}\left(\bm{m}, \bm{V}\right), \\
    \lambda_d^2 \mid \nu_d, \beta_d, \xi, \lambda_d^2 &\sim \operatorname{IG}\left(1, \frac{1}{\nu_d} + \frac{\beta_d^2}{2\tau^2} \right), \label{Eq. bclr-GS-HS-lambda}\\
    \nu_d \mid \lambda_d^2, \beta_d, \xi, \tau^2 &\sim \operatorname{IG}\left(1, 1+\frac{1}{\lambda_d^2}\right), \\
    \tau^2 \mid \nu_d, \beta_d, \xi, \lambda_d^2 &\sim \operatorname{IG}\left( \frac{p+1}{2}, \frac{1}{\xi} + \sum_{d=1}^p \frac{\beta_d^2}{2\lambda_d^2}\right), \\
    \xi \mid \tau^2 &\sim \operatorname{IG}\left(1, 1+\frac{1}{\tau^2}\right), \label{Eq. bclr-GS-HS-xi}
\end{aligned}
$$ for $d \in \{1, \dots, p\}$ and $i \in \{1, \dots, N\}$.

\end{document}